\documentclass[twocolumn]{aastex631} 
\usepackage{placeins}
\usepackage{xcolor}
\usepackage{amsmath}
\usepackage{amssymb}
\usepackage{chemformula}
\usepackage{comment}

\usepackage{chngpage}

\newcommand\standardstate{{\circ\kern-0.495em-}}

\begin{document}
\title{Towards a self-consistent evaluation of gas dwarf scenarios for temperate sub-Neptunes}
\author{Frances E. Rigby}
\altaffiliation{These authors contributed comparably to this work.}
\affiliation{Institute of Astronomy, University of Cambridge,
Madingley Road, Cambridge CB3 0HA, UK}

\author{Lorenzo Pica-Ciamarra}
\altaffiliation{These authors contributed comparably to this work.}
\affiliation{Institute of Astronomy, University of Cambridge,
Madingley Road, Cambridge CB3 0HA, UK}

\author{M\aa ns Holmberg}
\altaffiliation{These authors contributed comparably to this work.}
\affiliation{Institute of Astronomy, University of Cambridge, 
Madingley Road, Cambridge CB3 0HA, UK}

\author{Nikku Madhusudhan}
\altaffiliation{These authors contributed comparably to this work.}
\affiliation{Institute of Astronomy, University of Cambridge, 
Madingley Road, Cambridge CB3 0HA, UK}

\author{Savvas Constantinou}
\altaffiliation{These authors contributed comparably to this work.}
\affiliation{Institute of Astronomy, University of Cambridge, 
Madingley Road, Cambridge CB3 0HA, UK}

\correspondingauthor{Nikku Madhusudhan}
\email{nmadhu@ast.cam.ac.uk}

\author{Laura Schaefer}
\affiliation{Department of Geological Sciences, School of Earth, Energy, and Environmental Sciences,\\ Stanford University, Stanford, CA 94305, USA}

\author{Jie Deng}
\affiliation{Department of Geosciences, Princeton University, Princeton, NJ 08544, USA}

\author{Kanani K. M. Lee}
\affiliation{Department of Physics, United States Coast Guard Academy, New London, CT 06320, USA}

\author{Julianne I. Moses}
\affiliation{Space Science Institute, 4765 Walnut Street, Suite B, Boulder, CO 80301, USA}

\begin{abstract}
The recent JWST detections of carbon-bearing molecules in a habitable-zone sub-Neptune have opened a new era in the study of low-mass exoplanets. The sub-Neptune regime spans a wide diversity of planetary interiors and atmospheres not witnessed in the solar system, including mini-Neptunes, super-Earths, and water worlds. Recent works have investigated the possibility of gas dwarfs, with rocky interiors and thick H$_2$-rich atmospheres, to explain aspects of the sub-Neptune population, including the radius valley. Interactions between the H$_2$-rich envelope and a potential magma ocean may lead to observable atmospheric signatures. We report a coupled interior-atmosphere modelling framework for gas dwarfs to investigate the plausibility of magma oceans on such planets and their observable diagnostics. We find that the surface-atmosphere interactions and atmospheric composition are sensitive to a wide range of parameters, including the atmospheric and internal structure, mineral composition, volatile solubility and atmospheric chemistry. While magma oceans are typically associated with high-temperature rocky planets, we assess if such conditions may be admissible and observable for temperate sub-Neptunes. We find that a holistic modelling approach is required for this purpose and to avoid unphysical model solutions. We find using our model framework and considering the habitable-zone sub-Neptune K2-18~b as a case study that its observed atmospheric composition is incompatible with a magma ocean scenario. We identify key atmospheric molecular and elemental diagnostics, including the abundances of  CO$_2$, CO, NH$_3$ and, potentially, S-bearing species. Our study also underscores the need for fundamental material properties for accurate modelling of such planets. 
\end{abstract}

\section{Introduction} \label{sec:intro}

Sub-Neptune planets, with radii $1\ \mathrm{R}_\oplus \lesssim R_{\mathrm{p}} \lesssim 4\ \mathrm{R}_\oplus$, have emerged as the new frontier of exoplanet science and constitute the most numerous class of planets detected to date (e.g., \citealp{Fressin2013,Fulton2018}). The nature of the sub-Neptune population remains debated, as their bulk densities can be explained by a number of degenerate interior compositions \citep[e.g.][]{Rogers2011,Valencia2013}. These include rocky planets with diverse atmospheric compositions, mini-Neptunes with volatile-rich interiors and deep H$_2$-rich atmospheres, and water worlds with substantial water mass fractions, including Hycean worlds \citep[e.g.,][]{Rogers2011, Valencia2013, dorn_generalized_2017,  zeng_growth_2019, madhusudhan_interior_2020, madhusudhan_habitability_2021,rigby_ocean_2024}. 

The James Webb Space Telescope (JWST) is revolutionising our understanding of sub-Neptunes through high-precision atmospheric spectroscopy. JWST observations have led to confident detections and precise abundance constraints for CH$_4$ and CO$_2$ in the atmospheres of the habitable-zone sub-Neptune and candidate Hycean world \citep{madhusudhan_habitability_2021} K2-18~b \citep{madhusudhan_carbon-bearing_2023}, demonstrating the promise of JWST for detailed atmospheric characterisation. Furthermore, such observations are starting to be available for other temperate sub-Neptunes, including TOI-270~d \citep{Holmberg2024, Benneke2024} -- where abundance constraints for CH$_4$ and CO$_2$ were also retrieved -- and LHS~1140~b \citep{doyon_temperate_2024, Damiano2024}. Such precise abundance measurements pave the way towards understanding the interactions between the planet's atmosphere and interior, including the presence and nature of an underlying surface, as well as the planetary formation processes that give rise to such planets.

One of the most distinct features of the sub-Neptune population is the radius valley, a bimodal distribution of sub-Neptune radii with a minimum around 1.8~R$_{\oplus}$ \citep{fulton_california-kepler_2017, Fulton2018, Cloutier2020}. Two competing hypotheses have been proposed to explain the origin of the radius valley. One explanation suggests that the valley is a consequence of differential atmospheric mass loss between planets of different masses. In this hypothesis, both populations would be composed of planets with predominantly rocky interiors. The more massive planets would retain their primary H$_2$-rich atmospheres, while the less massive ones would instead largely lose their envelope and hence have a smaller radius. We refer to the larger population, with rocky interiors and a deep \ch{H2}-rich atmospheres, as gas dwarfs. The mechanism for the mass loss is debated, with the predictions of two hypotheses -- photoevaporation \citep[e.g., ][]{lopez_role_2013, jin_planetary_2014, owen_evaporation_2017, jin_compositional_2018} and core-powered mass loss \citep[e.g., ][]{ginzburg_super-earth_2016, ginzburg_core-powered_2018, gupta_sculpting_2019, gupta_signatures_2020} -- both proposed to explain the observations \citep{rogers_photoevaporation_2021}. A second explanation \citep[e.g., ][]{zeng_growth_2019, venturini_nature_2020, izidoro_formation_2021} suggests that the valley could instead be due to planets having different interior compositions. The smaller radius population would be rocky, as in the atmospheric mass-loss scenario, while the larger population would be composed of planets with water-rich interiors due to significant accumulation of icy planetesimals/pebbles during their formation and migration. Atmospheric observations of planets in the sub-Neptune range may be able to distinguish between these two scenarios \citep[e.g., ][]{kite_superabundance_2019, kite_atmosphere_2020, daviau_experimental_2021, gaillard_redox_2022, schlichting_chemical_2022, charnoz_effect_2023, misener_atmospheres_2023, falco_hydrogenated_2024}. 

While the gas dwarf hypothesis has garnered significant attention in the literature \citep[e.g.,][]{lopez_role_2013, jin_planetary_2014, ginzburg_super-earth_2016,owen_evaporation_2017, jin_compositional_2018, ginzburg_core-powered_2018, gupta_sculpting_2019, kite_superabundance_2019, gupta_signatures_2020, kite_atmosphere_2020, bean_nature_2021, schlichting_chemical_2022, charnoz_effect_2023}, several open questions remain. Firstly, it is unclear whether it is possible for rocky cores to accrete a substantial H$_2$-rich envelope without significant accretion of other volatiles and ices \citep{Fortney2013,venturini_fading_2024}. Secondly, should such planets exist, would the atmosphere-interior interactions give rise to distinct atmospheric signatures? This might be expected if the rocky surface were to be molten, giving rise to a magma ocean scenario \citep{Schaefer2016, Schaefer2017,  kite_superabundance_2019, kite_atmosphere_2020, daviau_experimental_2021, gaillard_redox_2022, schlichting_chemical_2022, misener_atmospheres_2023, charnoz_effect_2023, falco_hydrogenated_2024, shorttle_distinguishing_2024, tian_atmospheric_2024}. It is, however, not fully clear whether this scenario is possible, particularly for planets with a low equilibrium temperature. For these planets, only a subset of atmospheric structures, combining sufficient but not exceedingly high surface pressure and very high surface temperature, could result in magma at the base of the atmosphere.

Several recent studies have explored the implications of a magma ocean on the atmosphere and interior compositions of diverse planets, both with terrestrial-like \citep{Schaefer2016, Schaefer2017, daviau_experimental_2021, gaillard_redox_2022, tian_atmospheric_2024} and H$_2$-rich atmospheres \citep{kite_superabundance_2019, kite_atmosphere_2020, schlichting_chemical_2022, misener_atmospheres_2023, charnoz_effect_2023, falco_hydrogenated_2024, shorttle_distinguishing_2024, tian_atmospheric_2024}. These works identify several key factors, including temperature and oxygen fugacity at the bottom of the atmosphere, that influence the composition of the atmosphere, driven by thermochemical equilibrium at the gas-melt interface. For example, some notable atmospheric signatures of reduced conditions in a rocky interior include potential nitrogen depletion \citep[e.g., ][]{daviau_experimental_2021, dasgupta_fate_2022,suer_distribution_2023,shorttle_distinguishing_2024} and high CO/CO$_2$ ratio for H$_2$-rich atmospheres \citep{gaillard_redox_2022, schlichting_chemical_2022}. However, the interplay between the atmosphere, interior, and the corresponding surface-atmosphere interactions in sub-Neptunes is only beginning to be explored in a realistic manner \citep[e.g., ][]{kite_atmosphere_2020, schlichting_chemical_2022}.

In this work, we develop an integrated magma ocean framework for temperate, H$_2$-rich sub-Neptunes. Our framework, presented in Section~\ref{sec:methods}, includes atmospheric and internal structure modelling, melt-gas interactions, and both equilibrium and disequilibrium processes in the atmosphere, resulting in spectroscopic predictions of atmospheric observables. We consider thermochemical equilibrium at the magma-atmosphere interface, and the solubility of volatile (\ch{H}, \ch{C}, \ch{N}, \ch{O}, \ch{S}) bearing species in magma. We explore the extreme case of the habitable-zone sub-Neptune and Hycean candidate K2-18~b \citep{madhusudhan_interior_2020} to investigate the plausibility of a magma ocean \citep[e.g.][]{kite_atmosphere_2020,shorttle_distinguishing_2024} and, if present, its atmospheric signatures. In doing so, we first use our framework to perform a comparative assessment of previous works in this direction in Section~\ref{sec:previous}, both on terrestrial-like atmospheres \citep{gaillard_redox_2022} and on \ch{H2}-rich ones \citep{kite_superabundance_2019, kite_atmosphere_2020, schlichting_chemical_2022, charnoz_effect_2023, misener_atmospheres_2023, falco_hydrogenated_2024,shorttle_distinguishing_2024, tian_atmospheric_2024}, with a focus on the case study of the candidate Hycean world K2-18~b. We then present our model predictions in Section~\ref{sec:results}. Finally, we summarize our findings and discuss future work in Section~\ref{sec:discussion}, highlighting the need for physically consistent models, and new experimental and theoretical work to derive accurate fundamental material properties.

\section{Methods} \label{sec:methods}
\begin{figure}
    \centering
    \includegraphics[width=\columnwidth]{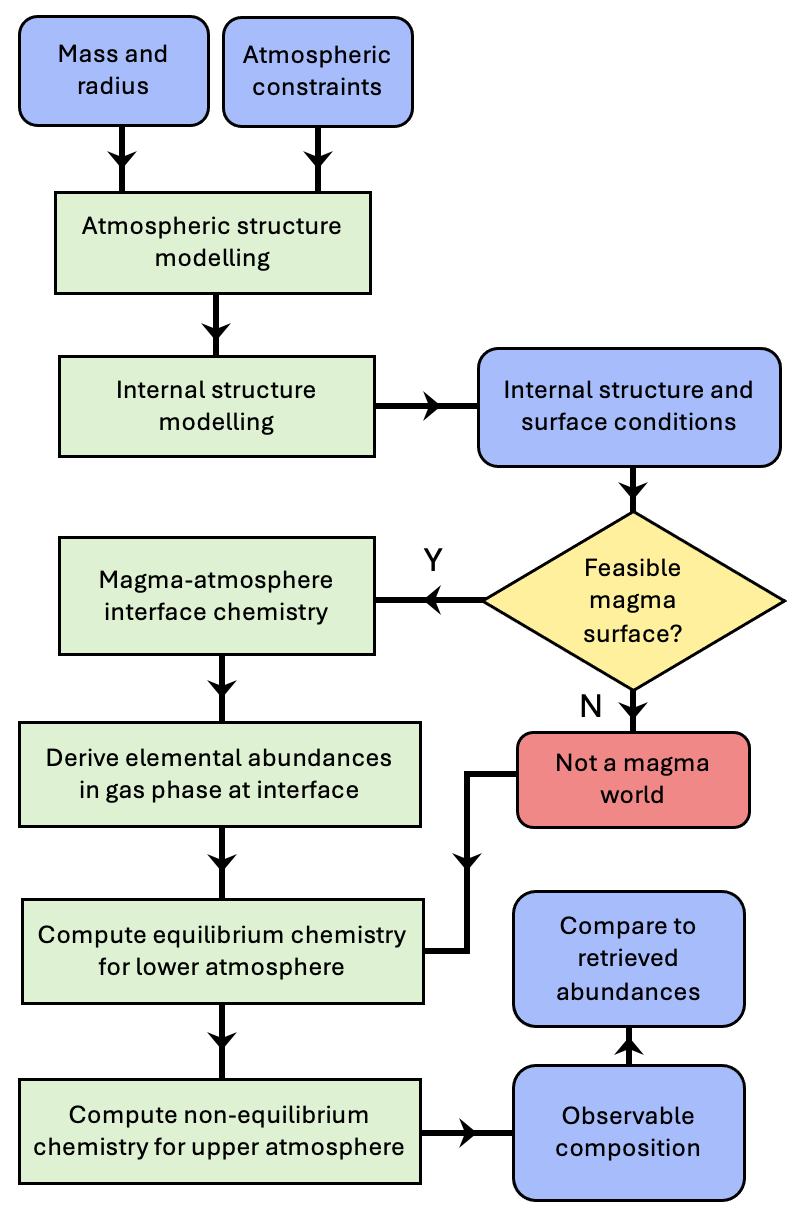}
    \caption{Flowchart showing an integrated modelling framework to assess gas dwarf scenarios for planets in the sub-Neptune regime.}
    \label{fig:flowchart}
\end{figure}

We develop an integrated modelling framework to evaluate gas dwarf scenarios for planets in the sub-Neptune regime. A schematic flowchart of the framework is shown in  Figure~\ref{fig:flowchart}.
We start by considering the constraints that the observed bulk parameters (mass, radius, and hence density) and known atmospheric properties impose on the planet's atmospheric and internal structure. This enables us to infer the possible conditions at the surface-atmosphere boundary, and, by considering a relevant mineral phase diagram, assess whether such conditions can in principle lead to a magma ocean scenario. If they can, we proceed by modelling the chemistry at the magma-atmosphere interface, which is determined by equilibrium processes including the solubility of relevant volatiles in the silicate melt, providing us with the elemental abundances in the gas phase at the interface. 
These are then evolved to the rest of the atmosphere, assuming chemical equilibrium in the lower atmosphere, and non-equilibrium processes (photochemistry and vertical mixing) in the upper atmosphere. 
This allows us to compute the observable composition of the atmosphere, which can be compared with the molecular abundances retrieved through observations to finally assess the plausibility of a magma ocean scenario for the planet.
We now describe in detail each of the steps outlined above. 

\subsection{Atmospheric Structure and Composition}
\label{sec:methods/atmospheric_structure}
We begin by modelling the atmospheric temperature structure in a self-consistent manner. In order to do so, the atmospheric chemical composition needs to be assumed. This can be done either by assuming the elemental abundances and atmospheric chemistry, or by directly assuming the molecular mixing ratios in the atmosphere. Other parameters that need to be taken into account include the internal temperature $T_{\mathrm{int}}$ representing an internal heat flux, the incident irradiation, the stellar properties, the presence and characteristics of clouds/hazes in the planet's atmosphere, and the efficiency of day-night energy redistribution. The self-consistent calculation will yield a pressure-temperature ($P$-$T$) profile, which will be coupled to the internal structure model, as discussed in Section~\ref{sec:methods/internal_structure}.

In order to carry out the self-consistent modelling of the atmospheric structure, we use the GENESIS framework \citep{Gandhi2017} adapted for sub-Neptunes (\citealp{madhusudhan_interior_2020, piette_temperature_2020, madhusudhan_habitability_2021, madhusudhan_chemical_2023}). GENESIS solves for radiative-convective equilibrium throughout the atmosphere, which is assumed to be plane-parallel, using the Rybicki scheme. It carries out line-by-line radiative transfer calculations through the Feautrier method \citep{hubeny_model_2017} and the discontinuous finite element method \citep{castor_new_1992}, while taking into account all of the parameters mentioned earlier in this section.  

For the atmospheric composition, we adopt uniform mixing ratios of molecular species based on the retrieved values at the terminator region of K2-18~b \citep{madhusudhan_carbon-bearing_2023}. We use the median retrieved abundances for the one-offset case: $\log{X_{\ch{CH4}}} = -1.72$ and $\log{X_{\ch{CO2}}} = -2.04$. For \ch{H2O}, we consider the 95\% one-offset upper limit, $\log{X_{\ch{H2O}}} = -3.01$. We also assume the incident irradiation and stellar properties of K2-18~b, and uniform day-night energy redistribution. We then explore the remaining parameter space. In particular, we consider two end-member values for $T_{\mathrm{int}}$, 25 K and 50 K, following \citet{madhusudhan_interior_2020} and \citet{Valencia2013}, and three values for $a$, the hazes' Rayleigh enhancement factor: 100, 1500 and 10000. We consider four combinations of these parameters, obtaining a cold case (designated C1, corresponding to $T_{\mathrm{int}} = 25$ K, $a = 10000$), two canonical cases (both with $a = 1500$, designated C2 for $T_{\mathrm{int}} = 25$ K and C3 for $T_{\mathrm{int}} = 50$ K) and a hot case (C4, with $T_{\mathrm{int}} = 50$ K and $a = 100$). These profiles are shown in Figure~\ref{fig:interiorresults}.

We place the upper boundary of the atmosphere at $10^{-6}$ bar, and calculate the $P$-$T$ profile self-consistently down to $10^3$ bar, below the radiative-convective boundary. At higher pressures, we extrapolate the profile as an adiabat, using the \ch{H2}/He equation of state (EOS), $\rho=\rho(P,T)$, and adiabatic gradient from \citet{Chabrier2019}. 
We note that, in principle, an appropriate $P$-$T$ profile may be even colder than C1, considering the constraints on clouds/hazes at the terminator from observations of K2-18~b \citep{madhusudhan_carbon-bearing_2023}.

\subsection{Internal Structure Modelling} \label{sec:methods/internal_structure}

\begin{figure}
    \centering
    \includegraphics[width=\columnwidth]{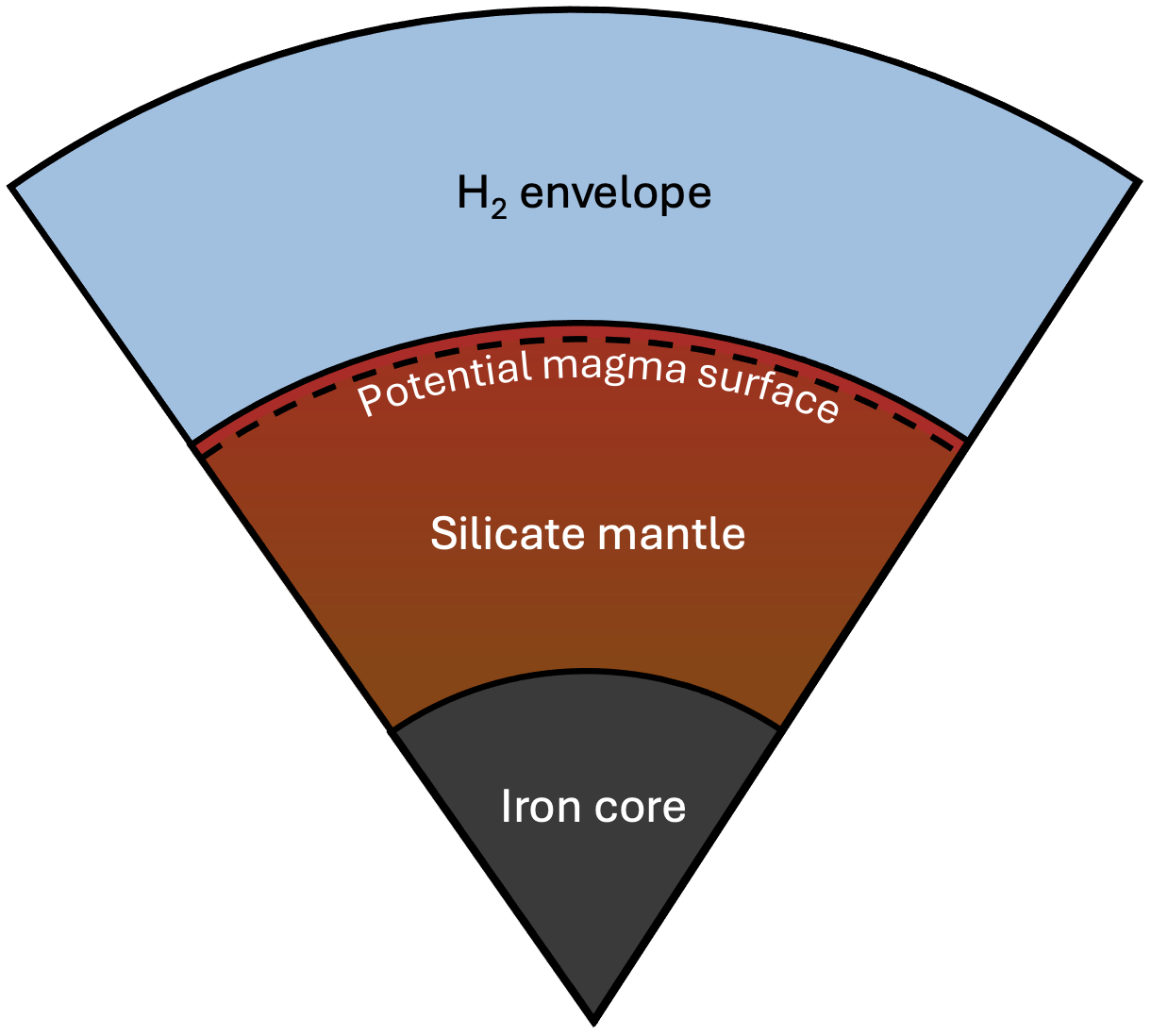}
    \caption{Cross-section of the internal structure of a potential gas dwarf, including the H$_2$-rich envelope, silicate mantle and iron core.}
    \label{fig:crosssection}
\end{figure}

We model planetary internal structures using the HyRIS framework, outlined in \citet{rigby_ocean_2024}. The model calculates the planet radius ($R_\mathrm{p}$) from the planet mass ($M_\mathrm{p}$), the mass fractions of the planet's components ($x_\mathrm{i}=M_\mathrm{i}/M_\mathrm{p}$), and the corresponding EOS and $P$-$T$ profile. HyRIS solves the equations for mass continuity and hydrostatic equilibrium using a fourth-order Runge-Kutta method, and solves for $R_\mathrm{p}$ using a bisection procedure. For the purpose of this study investigating magma-ocean scenarios, the internal structure model includes a H$_2$-rich envelope, a silicate mantle, and an iron core. 

\begin{figure*}
    \centering
    \includegraphics[width=1.9\columnwidth]{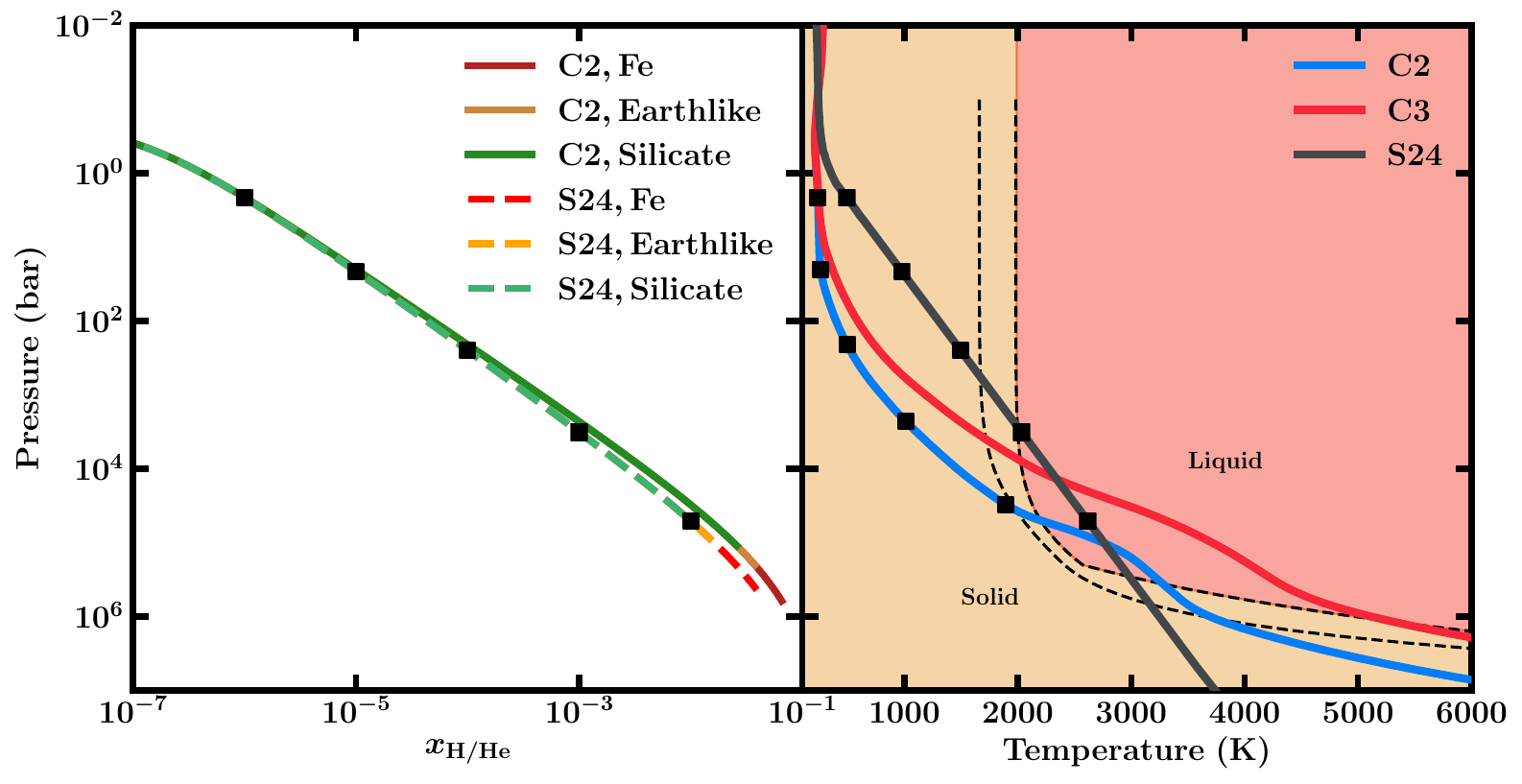}
    \caption{{\bf Left:} Pressure in the H$_2$-rich envelope against the envelope mass fraction above this pressure level. Solid and dashed lines show these assuming C2 and the profile used by S24 in the envelope respectively. The black squares indicate expected surface conditions for different envelope mass fractions, independent of satisfying the bulk properties of the planet. {\bf Right:} The nominal pressure-temperature profiles generated for this work and S24 shown against the liquidus and solidus for peridotite \citep{Fiquet2010,Monteux2016}. The solid and liquid phases are shaded. The black squares again show the corresponding surface conditions expected for the different envelope mass fractions.}
    \label{fig:surfaceconds}
    \vspace{2mm}
\end{figure*}

The silicate mantle is described by EOSs valid for the liquid and solid phases -- for simplicity, we adopt a separate EOS prescription on either side of a melting curve. The composition is nominally assumed to be peridotitic. The magma is described by an EOS for peridotitic melt compiled similarly to \citet{Monteux2016} by combining the densities of molten enstatite, forsterite, fayalite, anorthite and diopside, described by third-order Birch-Murnaghan/Mie-Gruneisen EOSs from \citet{Thomas2013}, weighted by their mass fractions. For the purpose of this initial study, we assume complete melting occurs at the liquidus, and hence do not include an EOS prescription for the partial melt between the solidus and liquidus curves. We use the peridotite liquidus from \citet{Monteux2016}, based on \citet{Fiquet2010} -- both the liquidus and solidus are shown in Figure~\ref{fig:surfaceconds} \citep{Fiquet2010,Monteux2016}. The solid portion of the silicate mantle is described by the EOS of \citet{Lee2004} for the high-pressure peridotite assemblage. At extreme mantle pressures beyond the pressure range of these experiments ($107$ GPa), we use the temperature-independent EOS of \citet{Seager2007} for MgSiO$_3$ perovskite, originally derived at room temperature. The thermal effects for solid silicates at these pressures are small \citep{Seager2007} with negligible effect on the internal structure. The iron core is described by the EOS of \citet{Seager2007} for hexagonal close-packed Fe.\\
The temperature structure in the melt is assumed to be adiabatic. The adiabatic gradient is calculated using the specific heat for peridotite from \citet{Monteux2016} and the volume expansion coefficient that we calculate from the combined peridotite melt EOS. The adiabatic gradient in the upper portion of the solid mantle is calculated following \citet{Lee2004}. Following previous studies \citep[e.g., ][]{Rogers2011,Nixon2021,rigby_ocean_2024}, the remaining solid portion of the interior is taken to be isothermal, as the EOSs used are temperature-independent \citep{Seager2007}.  

The mass of the magma ocean follows from the adiabatic temperature profile in the melt, similarly to the calculation of water ocean depths by \citet{Nixon2021} and \citet{rigby_ocean_2024}. The melt adiabat and hence the magma base pressure are defined by the surface pressure and temperature. For a given interior composition and surface conditions, the mass of the melt can thus be calculated. We adapt HyRIS to automate the extraction of the relevant melt characteristics, similar to the methods for water oceans in \citet{rigby_ocean_2024}. The mass fraction of the melt is an important quantity for considerations of the available volatile reservoir, as discussed below. We note that the moderate increase of the magma ocean mass fraction that may result from partial melting is partly accounted for by our range of considered melt masses in Section~\ref{sec:results/interface}.

\subsection{Melt-atmosphere Interface Chemistry} \label{sec:methods/melt_interface}

The atmospheric chemistry is constrained by the elemental composition at the bottom of the atmosphere, which is governed by the interactions at the magma ocean and atmosphere interface. At this boundary, we model the reactions and solubility of the gas species in thermochemical equilibrium. We include 82 H-C-N-O-S gas species and He, the set of which we denote $\mathbf{X}$, and their equilibrium reactions, nominally excluding other effects such as condensation and exsolution. Of these volatile species, we consider the solubility in the melt of H$_2$ \citep[basalt case]{hirschmann_solubility_2012}, H$_2$O \citep{iacono-marziano_new_2012}, CO \citep[MORB case]{yoshioka_carbon_2019}, CO$_2$ \citep{suer_distribution_2023}, CH$_4$ \citep{ardia_solubility_2013}, N$_2$ \citep{dasgupta_fate_2022}, S$_2$ \citep{gaillard_redox_2022} and \ch{H2S} \citep{clemente_solubility_2004}, as further motivated in Appendix~\ref{app:solubility}. We note that the solubility of \ch{H2S} is uncertain at high temperatures/pressures and may be higher if, for example, its solubility approaches that of \ch{S2}. Furthermore, we remark that we are not considering the possible exsolution of \ch{FeS}, which may affect the abundance of sulfur in the atmosphere. Likewise, the overall solubility of nitrogen is calculated here through \ch{N2}, and may be higher if the solubility of \ch{NH3} is significant. The data on \ch{NH3} solubility in magma is currently limited and it is difficult to make any quantitative estimates of \ch{NH3} solubility. For the explicitly composition-dependent laws, we use the \citet{iacono-marziano_new_2012} Etna basalt melt composition. Similarly to \citet{kite_atmosphere_2020}, we assume that the magma is well-stirred such that the equilibration at the surface sets the volatile abundance throughout the melt.

These solubility laws relate the partial pressures in the atmosphere to the concentrations of the volatiles in the melt. The amount of volatiles in the melt thus depends on the equilibrium chemistry, the solubility and the total mass of the melt, $M_{\textrm{melt}}$. For a given mass of the atmosphere and the melt, we have the following mass balance condition for each species $i$ \citep[similar to][]{gaillard_redox_2022},

\begin{equation} \label{eq:mass}
    M_{\textrm{tot}} \, w_i = M_{\textrm{atm}} \, w_{i, \textrm{atm}} + M_{\textrm{melt}} \, w_{i, \textrm{melt}}\,,
\end{equation}
where $w_i$ is the total mass fraction of each species $i$.

To determine the chemical composition of the atmosphere and the melt, we solve the element conservation equations
\begin{equation} \label{eq:el_cons}
    \varepsilon_j = \sum_{i \, \in \, \mathbf{X}} \nu_{ij} \frac{n_i}{n_{\langle H \rangle}}\,,
\end{equation}
where $n_i$ is the total amount of moles of species $i$, $n_{\langle H \rangle} = n_H + 2n_{H_2} + 2n_{H_2O} + \dots$ is the total amount of moles of hydrogen, $\nu_{ij}$ are the coefficients of the stoichiometric matrix, and $\varepsilon_j$ is the elemental abundance of element $j$ relative to hydrogen. Equation~\eqref{eq:el_cons} is coupled to Equation~\eqref{eq:mass} via $n_i \propto w_i / \mu_i$, where $\mu_i$ is the molar mass, which in turn is coupled to the law of mass action
\begin{equation} \label{eq:mass_action}
    \frac{p_i}{p^\standardstate} = K_i \prod_{j \, \in\,  E} \left(  \frac{p_j}{p^\standardstate} \right)^{\nu_{ij}}\,,
\end{equation}
and the solubility laws, determining both $w_{i, \textrm{atm}}$ and $w_{i, \textrm{melt}}$. Here, $E$ is the set of all elements, $p_i$ is the partial pressure of species $i$, $K_i$ is the temperature-dependent equilibrium constant, and $p^\standardstate$ is a standard pressure of 1 bar. For each gas species, we approximate the equilibrium constant as
\begin{equation}
    \ln K(T) = \frac{a_0}{T} + a_1 \ln T + b_0 + b_1 T + b_2 T^2\,,
\end{equation}
using the coefficients provided by FASTCHEM \citep{Stock2018, Stock2022}, mainly derived using thermochemical data from \cite{Chase1998}.

Overall, Equation~\eqref{eq:mass_action} depends on the elemental partial pressures, with 6 unknowns, corresponding to the 6 elements considered. Nominally, we solve for these using the 5 equations in \eqref{eq:el_cons} for all elements apart from H, together with
\begin{equation}
    P_s = \sum_{i \, \in\,  \mathbf{X}} p_i\,,
\end{equation}
to fix the total pressure. This treatment of oxygen yields a first-order estimate of the redox state as set by the atmosphere. Alternately, we consider oxygen fugacity ($f_{\ch{O2}}$) as a free parameter, by determining $p_{\ch{O}}$ in Equation~\eqref{eq:mass_action} via $f_{\ch{O2}} =  p_{\ch{O2}} = K_{\ch{O2}} \, p_{\ch{O}}^2  / p^\standardstate$, allowing us to consider different redox conditions. In this framework, we assume ideal gas behavior such that fugacity and partial pressure are equivalent \citep[e.g.][]{Bower2022, schlichting_chemical_2022}. 

As a cross-check, we validate our new framework against a self-consistent atmosphere composition model \citep[e.g.][]{Schaefer2017} which uses the Gibbs energy minimization code IVTANTHERMO \citep{BELOV1999173}. IVTANTHERMO uses a thermodynamic database based on \citet{gurvich1990thermodynamic}, which we modify to include the silicate-melt dissolved volatile species H$_{2}$, OH$^{-}$, O$^{2-}$, CO, CO$_{2}$, CH$_{4}$, N$^{3-}$, and S$^{2-}$. We calculate equilibrium between a total possible 366 gas species and 201 condensed species. For the dissolution reactions, we assume $\Delta$C$_{p}$ = 0 and that any temperature dependence in the equilibrium constant is due to the heat of the reaction. However, data is available only in limited temperature ranges for most dissolution reactions, so we assume a simple Henry's law solubility relation for all of the dissolved species except S$^{2-}$, OH$^{-}$, H$_{2}$, and CH$_{4}$. We also neglect non-ideality in both the gas phase and melt. Using IVTANTHERMO, we then compute self-consistent equilibrium between the gas phase and melt species as a function of pressure and temperature.

For this comparison, we use 50$\times$solar bulk elemental abundances (not including He), $P_\mathrm{s}=10^4$~bar, $T_\mathrm{s}=3000$~K, and $M_{\textrm{melt}} / M_{\textrm{atm}} = 0.20$, which is given by the gas-to-melt mass ratio as calculated by IVTANTHERMO. We find that all major H-C-N-O-S gas species agree to within at most 0.35 dex (standard deviation of 0.1 dex), with the largest deviation coming from \ch{CO2}. This deviation mostly stems from the oxygen fugacities being somewhat different between the two approaches, with IVTANTHERMO yielding a 0.35 dex lower value. Furthermore, we verify that we recover the atmospheric abundances given by FASTCHEM 2 \citep{Stock2022} and GGCHEM \citep{Woitke2018} when setting $M_{\textrm{melt}} = 0$.

\subsection{Atmospheric Chemistry}
\label{sec:methods/atmospheric_chemistry}
We carry out equilibrium and disequilibrium chemistry calculations to determine the atmospheric composition above the magma/rock surface. We use the VULCAN photochemical kinetics framework \citep{Tsai2021}, with the initial atmospheric chemistry obtained using the FASTCHEM equilibrium chemistry code \citep{Stock2018}.

For equilibrium chemistry calculations, we consider thermochemical equilibrium involving H-C-N-O-S species as well as He, along with H$_2$O condensation. For calculations considering disequilibrium processes, we additionally include the effects of vertical mixing and photochemistry. We follow the $K_\mathrm{zz}$ parameterisation of \citet{madhusudhan_chemical_2023}:
\begin{equation}
K_\mathrm{zz} / \mathrm{cm}^2 \mathrm{s}^{-1} = 
\begin{cases}
    \mathrm{min}(\frac{5.6 \times 10^{4}}{ (P/\mathrm{bar})^{\frac{1}{2}}}, 10^{10}), & P \leq 0.5~\mathrm{bar} \\
    10^{6}, & P > 0.5~\mathrm{bar},
\end{cases}
\end{equation}
although we note that the K$_\mathrm{zz}$ in the troposphere could be higher (e.g., $\sim 10^7-10^8$ cm$^2$~s$^{-1}$ in the deep convective region of the atmosphere) or lower (e.g., $\sim$10$^4$ cm$^2$~s$^{-1}$) in any radiative regions if moist convection is inhibited by the high molecular weight of water in the H$_2$-rich atmosphere (see \citealp{leconte_3d_2024}). Accordingly, we consider a wider range of $K_\mathrm{zz}$ values than our canonical treatment in Section~\ref{sec:results/5} and Appendix~\ref{app:atmosphericchem}.

Additionally, we consider photochemical reactions including H-C-N-O-S species, using a nominal stellar spectrum from the HAZMAT spectral library \citep{Peacock2020} corresponding to a median 5 Gyr star of radius 0.45 R$_\odot$ following previous work \citep{madhusudhan_chemical_2023}. We also specifically consider the condensation of H$_2$O to liquid and solid droplets, which fall at their terminal velocity, as described in \citet{Tsai2021}. We note that while the H-C-N-O chemistry has been extensively explored for sub-Neptunes in various studies \citep[e.g.][]{Yu2021,Hu2021,Tsai2021,madhusudhan_chemical_2023}, the S chemistry has not been explored in significant detail and may be incomplete. Nevertheless, we include S using the VULCAN framework \citep{Tsai2021} for completeness.

With the above calculations we obtain the vertical mixing ratio profiles for a number of relevant chemical species in the atmosphere. The abundances of key species in the observable part of the atmosphere can then be compared against constraint retrieved from an atmospheric spectrum. 

\subsection{Spectral Characteristics}
\label{sec:methods/spectral_characteristics}

We use the results of the chemistry calculation described in Section~\ref{sec:methods/atmospheric_chemistry} to simulate how such an atmosphere would appear in transmission spectroscopy, including the spectral contributions of relevant species. For this, we use the forward model generating component of the VIRA retrieval framework \citep{Constantinou2024}, which treats the planet’s terminator as a 1D atmosphere in hydrostatic equilibrium. We consider atmospheric opacity contributions from H$_2$O \citep{Barber2006, Rothman2010}, CH$_4$ \citep{Yurchenko2014}, NH$_3$ \citep{Yurchenko2011}, CO \citep{Li2015}, CO$_2$ \citep{Tashkun2015}, C$_2$H$_2$ \citep{Chubb2020}, HCN \citep{Barber2014}, H$_2$S \citep{Azzam2016, Chubb2018} and SO$_2$ \citep{Underwood2016}. We do not include N$_2$ in the model, as it has no significant absorption features in the near-infrared and it is not present in significant enough quantities to affect the atmospheric mean molecular weight. We additionally consider atmospheric extinction arising from H$_2$-H$_2$ and H$_2$-He collision-induced absorption \citep{Borysow1988, Orton2007, Abel2011, Richard2012}, which provide the spectral baseline, as well as H$_2$ Rayleigh scattering. We simulate transmission spectra using the vertical mixing ratio profiles computed using VULCAN as described above, and the $P$-$T$ profile appropriate to each case considered.  

\section{Results: Comparison with previous work} \label{sec:previous}

We now apply the framework described in Section~\ref{sec:methods} and compare with previous works on both terrestrial-like and sub-Neptune atmospheres.

\subsection{Terrestrial-like Atmospheres} 

Many previous studies have investigated surface-atmosphere interactions for magma oceans underneath terrestrial-like atmospheres \citep[e.g.,][]{Matsui1986,ElkinsTanton2008,Hamano2013,Lebrun2013, Wordsworth2016, kite_water_2021, lichtenberg_vertically_2021,Bower2022,gaillard_redox_2022}. Recent studies have explored the implications of diverse interiors of exoplanets for their atmospheric compositions. \citet{daviau_experimental_2021} proposed that, for reduced conditions, nitrogen is expected to be preferentially sequestered in the mantle, providing a valuable way to study the interior composition of such exoplanets. More recently, \citet{gaillard_redox_2022} investigated the primordial distribution of volatiles within the framework of melt-atmosphere interactions and discussed applications for Venus and Earth. For the early Earth, they find that reduced conditions, with oxygen fugacity two dex below the iron-w\"{u}stite (IW) buffer, $f_{\ch{O2}} \lesssim \ch{IW} -2$, result in an atmosphere abundant in \ch{H2}, \ch{CO} and \ch{CH4} but depleted in \ch{CO2} and \ch{N2}. On the other hand, for $f_{\ch{O2}} \gtrsim \ch{IW}+2$, \ch{CO2} becomes the main atmospheric component, with significant levels of \ch{SO2}, \ch{N2} and \ch{H2O}. In particular, the behaviour of nitrogen is a consequence of the high solubility of \ch{N2} as \ch{N^3-} in silicate melt at reducing conditions \citep[e.g., ][]{libourel_nitrogen_2003, dasgupta_fate_2022}, via the following reaction
\begin{equation}
    \frac{1}{2} \ch{N2}_{\,(\text{gas})} + \frac{3}{2} \ch{O^2-}_{\,(\text{melt})} \,\rightleftharpoons\,  \ch{N^3-}_{\,(\text{melt})} + \frac{3}{4} \ch{O2}_{\,(\text{gas})}\,.
\end{equation}
As a result, the melt concentration of \ch{N^3-} is proportional to $f_{\ch{N2}}^{1/2} f_{\ch{O2}}^{-3/4}$, thus favouring low $f_{\ch{O2}}$. In conclusion, these works predict that the abundance of atmospheric nitrogen may be used as a diagnostic for the redox state of a rocky planet's mantle.

\begin{figure}
    \centering
    \includegraphics[width=\columnwidth]{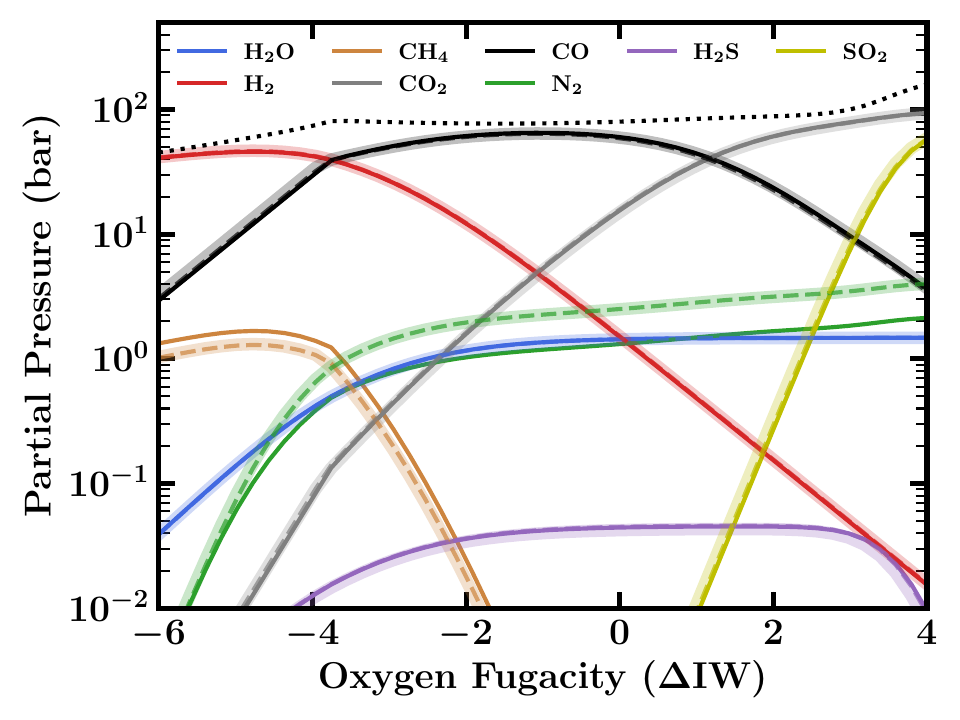}
    \caption{Comparison with \cite{gaillard_redox_2022}, showing the atmospheric composition as a function of oxygen fugacity. The solid lines show the partial pressures from our melt-atmosphere framework, described in Section~\ref{sec:methods/melt_interface}. The transparent lines are computed using the code from \cite{gaillard_redox_2022}, and the dashed lines are computed with our framework modified to approximate the results from \cite{gaillard_redox_2022}. The dotted line at the top illustrates the total surface pressure.}
    \label{fig:Gaillard}
\end{figure}

As a benchmark, we compare our melt-atmosphere equilibrium chemistry framework with \cite{gaillard_redox_2022}. We use their case with a magma ocean mass of half the bulk silicate mantle at $T = 1773$~K and with volatile contents of 90, 102, 3.3, and 126 ppm-wt for C, H, N, and S, respectively. With this, we reproduce their atmospheric composition as shown in Figure~\ref{fig:Gaillard}. Compared to our nominal setup in Section~\ref{sec:methods/melt_interface}, we added a constraint for the hydrogen abundance and solved for the resulting mass of the atmosphere, coupled to the surface pressure using \citep{gaillard_redox_2022}
\begin{equation}
    P_s = \frac{g M_{\textrm{atm}}}{4 \pi R_\mathrm{p}^2}\,,
\end{equation}
where $g$ is the gravitational acceleration at $R_\mathrm{p}$. For a like-to-like comparison, we added the condensation of graphite and used the same gas species (excluding Ar) and solubility laws as \cite{gaillard_redox_2022}. Overall, we find good agreement between both implementations, with the most deviation coming from \ch{N2} and \ch{CH4}. We find that the \ch{N2} discrepancy comes from an inconsistency in the code by \cite{gaillard_redox_2022}, whereby they use a molar mass of 14 g/mol for \ch{N2} instead of 28 g/mol. The remaining discrepancy is likely a result of minor differences in the implementations of the different reactions. We find that by accounting for some of these differences we can better match the result by \cite{gaillard_redox_2022}, as shown in Figure \ref{fig:Gaillard}. For this purpose, in addition to considering their adopted molar mass, we implemented the reactions $\ch{CH4} + 2 \ch{O2} \rightleftharpoons 2 \ch{H2O} + \ch{CO2}$ and $\ch{H2O} \rightleftharpoons 0.5 \ch{O2} + \ch{H2}$ using the equilibrium constants by \cite{gaillard_redox_2022} to obtain the partial pressures of \ch{CH4} and \ch{H2O}, instead of deriving these from the elemental partial pressures as described in Section~\ref{sec:methods/melt_interface}. We also used the oxygen fugacity of the IW buffer from \cite{gaillard_redox_2022} instead of \cite{Hirschmann2021}.

\subsection{Sub-Neptunes with H$_2$-rich Atmospheres}

Several recent studies have also explored magma-atmosphere interactions in sub-Neptunes with rocky interiors and H$_2$-rich atmospheres. \citet{kite_superabundance_2019} considered the impact of \ch{H2} solubility in silicate melts on the radius distribution of sub-Neptunes, addressing the \textit{radius cliff}, a sharp decline in the abundance of planets with $R_p \gtrsim 3 \mathrm{R}_{\oplus}$. 
They find that the high solubility of \ch{H2} in magma, especially at high pressure, limits the maximum radius that can be attained by sub-Neptunes through accretion of atmospheric \ch{H2}.
For a 10 $\mathrm{M}_\oplus$ core they find a limiting mass fraction of $1.5$ wt$\%$ \ch{H2} in the atmosphere -- corresponding to $> 20$ wt$\%$ \ch{H2} in the planet --, as any additional \ch{H2} would be stored almost exclusively in the interior. Looking at smaller planets ($2\ \mathrm{R}_{\oplus} \leq R_{\mathrm{p}} \leq 3\ \mathrm{R}_{\oplus}$), \citet{kite_atmosphere_2020} find that magma-atmosphere interactions would significantly affect the atmosphere's composition and mass. For example, a key insight is that the \ch{H2O}/\ch{H2} ratio in the atmosphere reflects not only external water delivery, but also water production as a result of atmosphere-magma interactions. This would make the \ch{H2O}/\ch{H2} a good diagnostic for atmospheric origin, as well as for magma composition. In particular, it is found to be proportional to the magma \ch{FeO} content. 

Further investigations were carried out by \citet{schlichting_chemical_2022}, \citet{charnoz_effect_2023} and, most recently, \citet{falco_hydrogenated_2024}.
Considering a surface temperature $T_{\rm s} = 4500$ K, $1 \%$ to $14 \%$ \ch{H} mass fractions (of overall planet mass) and model parameters resulting in $f_{\ch{O2}} \lesssim \ch{IW} - 2$, \citet{schlichting_chemical_2022} find that the atmosphere is expected to be dominated across the explored parameter space by \ch{H2}, \ch{SiO}, \ch{CO}, \ch{Mg} and \ch{Na}, followed by \ch{H2O}, which should exceed \ch{CO2} and \ch{CH4} by two to three orders of magnitude. It should be noted they do not include \ch{N} in their model. \citet{charnoz_effect_2023} and \citet{falco_hydrogenated_2024}, instead, consider total hydrogen pressures ranging between $10^{-6}$ and $10^6$ bar, temperatures between 1800 and 3500 K, and do not include any volatiles in their calculations, but also show that detectable absorption features of \ch{H2O} and \ch{SiO} should be expected. Additionally, the volatile-free investigation by \citet{misener_atmospheres_2023} finds that silane (\ch{SiH4}) should also be expected, dominating over \ch{SiO} at $P \gtrsim 0.1$ bar for an isothermal $T = 1000$~K pressure-temperature ($P$-$T$) profile in the upper atmosphere. Most recently, \citet{tian_atmospheric_2024} also investigated the outgassing mechanism for hybrid atmospheres in sub-Neptunes, but without considering solubilities in magma.\\

\subsection{End-member Scenario of K2-18~b} \label{previous/S24}

Some of the principles described above were recently applied to the habitable-zone sub-Neptune K2-18 b by \citet{shorttle_distinguishing_2024}, hereafter S24. Similarly to \citet{schlichting_chemical_2022} and \citet{gaillard_redox_2022}, S24 point to a high \ch{CO}/\ch{CO2} ratio and, like \citet{daviau_experimental_2021} and \citet{gaillard_redox_2022}, a depletion in atmospheric \ch{N} as signatures for the presence of a magma ocean and/or a reduced interior. It should be noted that the case of K2-18~b constitutes an end-member scenario. While most of the work on magma oceans has focused on very hot planets \citep[e.g.,][]{kite_atmosphere-interior_2016, Schaefer2016, kite_atmosphere_2020,  gaillard_redox_2022, charnoz_effect_2023, misener_atmospheres_2023, falco_hydrogenated_2024}, K2-18~b is a temperate sub-Neptune with equilibrium temperature $T_\mathrm{eq} = 272$ K (assuming an albedo of 0.3), close to that of the Earth. Here, we assess the findings of S24 using the framework described in Section~\ref{sec:methods} and Figure~\ref{fig:flowchart}.

We briefly note that in addition to gas dwarf and Hycean world scenarios, a mini-Neptune scenario with a thick H$_2$-rich atmosphere has also been proposed for K2-18~b \citep[e.g.][]{Hu2021,wogan_jwst_2024}. \citet{wogan_jwst_2024} conduct photochemical modelling of mini-Neptune cases for K2-18~b, suggesting a plausible solution. However, as noted in \citet{glein_geochemical_2024}, the calculated abundances are unable to match the retrieved abundances \citep{madhusudhan_carbon-bearing_2023}. In particular, the mixing ratios of CO and NH$_3$ are too large compared to the retrieved abundances, and so is the CO/CO$_2$ ratio.

\subsubsection{Consistency with bulk parameters}
\label{subsec:bulkparameters}

At the outset, it is important to ensure that any assumption about the internal structure is consistent with the planetary bulk parameters. Previous studies have shown that the bulk parameters of K2-18 b allow a degenerate set of solutions between a mini-Neptune, a Hycean world, or a rocky world with a thick H$_2$-rich atmosphere, i.e. a gas dwarf \citep{madhusudhan_interior_2020,madhusudhan_habitability_2021,rigby_ocean_2024}. Considering the present gas dwarf scenario, a purely rocky interior would require a minimum H$_2$-rich envelope mass fraction of $\sim$1\% \citep{madhusudhan_interior_2020}, as discussed below.

The model grid of S24 contains four values of mantle mass fraction relative to the total planet mass (0.001, 0.01, 0.1, and 1) and five values of the hydrogen mass fraction relative to the mantle mass (1, 10, 100, 1000 and 10000 ppm). Firstly, all the cases with a mantle mass fraction of 1 violate mass balance, as the sum of the mantle and atmospheric masses would exceed the total planet mass. Secondly, for the gas dwarf scenario, as noted above, the bulk density of K2-18 b requires an H$_2$-rich atmosphere with a minimum mass fraction of $\sim$1\%. In the S24 model grid, there is only one model which has an atmospheric mass fraction of 1\%, and it corresponds to a mantle mass fraction of 1, as noted above. It follows that all the remaining cases, with H$_2$ mass fraction below $1\%$, are incompatible with the planet's bulk density. 

In order to estimate the allowed atmospheric mass fractions for K2-18~b in the gas dwarf scenario, we consider four possible interior compositions, illustrated in Table~\ref{tab:extreme}: $f_\mathrm{silicate}=100 \%$, Earth-like ($f_\mathrm{silicate}=67 \%$), Mercury-like ($f_\mathrm{silicate}=30 \%$) and $f_\mathrm{silicate}=5 \%$, where $f_\mathrm{silicate}$ is the mass fraction of the interior (i.e. excluding the envelope) in the silicate mantle. We include $f_\mathrm{silicate}=5 \%$ as an end-member case, close to the upper limit for the allowed envelope mass fraction. Similarly, the extreme pure-silicate interior case is included as an end-member, yielding the lower limit on the allowed envelope mass fraction for a gas dwarf scenario. We adopt the median planetary mass $M_\mathrm{p}=8.63\ \mathrm{M_\oplus}$ \citep{cloutier_confirmation_2019} and radius $R_\mathrm{p}=2.61\ \mathrm{R_\oplus}$ \citep{benneke_water_2019} of K2-18~b. The allowed envelope mass also depends on the choice of $P$-$T$ profile, with hotter profiles leading to lower envelope masses for a given interior composition, as shown in Table~\ref{tab:extreme}. 

Considering the four self-consistent $P$-$T$ profiles described in Section~\ref{sec:methods/atmospheric_structure}, we find that an envelope mass fraction $x_{\mathrm{env}} \geq 1.34 \%$ is required for consistency with the bulk parameters. This limit corresponds to the extreme case of a 100 \% silicate interior, adopting C4 for the envelope $P$-$T$ profile. For a like-to-like comparison with the S24 model grid, we also consider their $P$-$T$ profile, which is the profile from \citet{benneke_water_2019} log-linearly extrapolated to higher pressures. For this profile, we find envelope mass fractions of $x_{\mathrm{env}} \geq 0.90\%$ are required, again corresponding to the extreme 100 \% silicate interior case. Overall, we find that all the models in the model grid of S24 are incompatible with mass balance and/or the bulk density of the planet considered. We demonstrate a self-consistent approach of accounting for the observed bulk parameters of K2-18~b in such calculations in Section \ref{sec:results}.

\subsubsection{Feasibility of a magma ocean} \label{subsubsec:interior}

As described in Section~\ref{sec:methods/internal_structure} and shown in Figure~\ref{fig:surfaceconds}, given an interior composition, the choice of $P$-$T$ profile affects the resulting envelope mass fraction. This, in turn, determines the surface pressure and temperature and the liquid/solid phase of the rocky surface underneath. Therefore, it is important to consider a physically motivated $P$-$T$ profile in the envelope. As mentioned above, S24 consider the $P$-$T$ profile from \citet{benneke_water_2019} at low pressures ($P \leq 4$ bar) and perform a log-linear extrapolation to the deep atmosphere ($P \gtrsim 10^5$ bar). The resulting temperature gradient can be significantly different from other self-consistent model $P$-$T$ profiles for the H$_2$-rich envelope \citep[e.g.,][]{Hu2021a,madhusudhan_chemical_2023,leconte_3d_2024}; an example is shown in Figure~\ref{fig:surfaceconds}.

We also note, however, that the actual surface temperature at the magma-atmosphere interface ($T_{\rm s}$) used in S24 appears to be a free parameter rather than self-consistently determined from their $P$-$T$ profile. The $T_{\rm s}$ ranges between $1500$ K and $3000$ K, but the corresponding pressure is not clear, considering their assertion that the maximum surface pressure allowed by the model is $10^8$ bar. This pressure also appears to be inconsistent with their maximum envelope mass fraction of $1\%$. Across the range of rocky compositions we consider, the maximum pressure reached is $\sim$5-7$\times$10$^5$ bar for envelope mass fractions $\sim$5-7\% depending on the $P$-$T$ profile as shown in Table~\ref{tab:extreme} and Figure~\ref{fig:surfaceconds}. 

Nevertheless, in order to establish the feasibility of achieving melt conditions in the S24 model, we consider the five highest envelope mass fractions used in S24. We adopt their mantle mass fraction of 1 and the corresponding five \ch{H2} mass fractions in their model grid, with a maximum of $1\%$. We then use these envelope mass fractions and the S24 $P$-$T$ profile to determine the corresponding expected surface pressures and temperatures, independent of satisfying the planetary bulk properties. These model points are shown in Figure~\ref{fig:surfaceconds} along with the liquidus and solidus curves for peridotite \citep{Fiquet2010,Monteux2016}. We find that only two of these five cases result in a magma surface in our framework. Finally, since we considered only the five highest envelope mass fractions of S24, it follows that all of the other models would also be unlikely to result in melt. We further note that for the two cases that result in a magma surface in S24, the magma mass fraction they consider is equal to the planet mass. However, based on the temperature structures shown in Figure~\ref{fig:interiorresults}, we find that the maximum magma mass fraction across the different interior compositions is $\sim$13\%, potentially somewhat higher as a result of partial melting, but not 100\%. 

\begin{table*}[ht!]
\centering
\begin{adjustwidth}{-0.82in}{-.5in}  
    \begin{tabular}{|c|c|c|c|c|c|c|c|c|c|c|}
        \hline
        $P$-$T$ & $f_\mathrm{silicate}$ & $x_\mathrm{interior}$ & $x_\mathrm{env}$ & $T_{\rm s}$  & $P_{\rm s}$ & $x_\mathrm{melt}$ & C/H & N/H & O/H & S/H \\
        Profile & ($\%$ of interior) & ($\%$) & ($\%$) & (K) & ($10^5$ bar) & ($\%$) & (log) & (log) & (log) & (log)\\
        \hline  
        C1 & 5 & 93.01 & 6.99 & 3278 & 6.52 & 0 & -1.84 & -2.47 & -1.61 & -3.18\\
        & 30 & 94.35 & 5.65 & 3120 & 3.99 & 0 & -1.84 & -2.47 & -1.61 & -3.18\\
        & 67 & 95.91 & 4.09 & 2928 & 2.15 & 0.86 & -1.84 & -4.54 & -1.65 & -3.21\\
        & 100 & 97.10 & 2.90 & 2664 & 1.23 & 2.02 & -1.84 & -3.51 & -1.70 & -3.25\\ \hline  
        C2 & 5 & 93.40 & 6.60 & 3461 & 6.27 & 0 & -1.84 & -2.47 & -1.61 & -3.18\\
        & 30 & 94.71 & 5.29 & 3290 & 3.79 & 0 & -1.84 & -2.47 & -1.61 & -3.18\\
        & 67 & 96.24 & 3.76 & 3084 & 2.00 & 1.81 & -1.84 & -4.36 & -1.69 & -3.25\\
        & 100 & 97.38 & 2.62 & 2819 & 1.12 & 3.16 & -1.83 & -3.34 & -1.76 & -3.32\\ \hline
        C3 & 5 & 94.94 & 5.06 & 4503 & 5.05 & 2.62 & -1.83 & -6.43 & -1.75 & -3.33\\
        & 30 & 96.17 & 3.83 & 4200 & 2.83 & 5.78 & -1.83 & -4.86 & -1.89 & -3.56\\
        & 67 & 97.48 & 2.52 & 3870 & 1.36 & 10.16 & -1.83 & -3.67 & -2.07 & -3.94\\
        & 100 & 98.38 & 1.62 & 3512 & 0.70 & 11.91 & -1.82 & -3.13 & -2.15 & -4.18\\ \hline 
        C4 & 5 & 95.43 & 4.57 & 4601 & 4.68 & 3.65 & -1.83 & -6.16 & -1.81 & -3.41\\
        & 30 & 96.61 & 3.39 & 4281 & 2.56 & 7.43 & -1.83 & -4.69 & -1.98 & -3.72\\
        & 67 & 97.84 & 2.16 & 3910 & 1.19 & 11.67 & -1.82 & -3.59 & -2.15 & -4.13\\
        & 100 & 98.66 & 1.34 & 3506 & 0.59 & 12.53 & -1.81 & -3.10 & -2.21 & -4.33\\ \hline  
    \end{tabular}
    \end{adjustwidth}
    \caption{\ch{H}/\ch{He} envelope mass fraction, resulting surface temperature, pressure, melt mass fraction constrained by the median values of the K2-18~b bulk parameters ($R_\mathrm{p} = 2.61\ \mathrm{R}_\oplus$, $M_\mathrm{p} = 8.63\ \mathrm{M}_\oplus$), and atmospheric elemental abundances. We use four interior compositions, where 5\% and 100\% silicate are unrealistic extreme cases included for completeness, and the four self-consistent $P$-$T$ profiles generated for this work. We note that the cases with 0 \% may include a region of partial melt. The bulk elemental abundances in the melt and atmosphere combined are set to 50$\times$solar.}
    \label{tab:extreme}
\end{table*}

\subsubsection{Magma-atmosphere interactions}

If the plausibility of a magma ocean is established, the melt-atmosphere interaction must be considered to determine its effect on the atmospheric composition. As described in Section~\ref{sec:methods/melt_interface}, the gas phase composition depends on the pressure and temperature at the interface, the elemental abundances, the amount of magma available, the solubilities of the chemical species, and the chemical properties of the melt. 

For the case of K2-18~b, S24 consider oxygen fugacity as a free parameter and assess the abundances of several H-C-N-O species in the lower atmosphere following melt-atmosphere interactions. They determine the atmospheric composition by considering three reactions, $\ch{CO2} + 2 \ch{H2} \rightleftharpoons \ch{CH4} + \ch{O2}$, $2\ch{CO2} \rightleftharpoons 2 \ch{CO} + \ch{O2}$, and $2\ch{H2O} \rightleftharpoons 2 \ch{H2} + \ch{O2}$, in thermochemical equilibrium, and solubilities of CH$_4$, N$_2$, CO$_2$, and H$_2$O in the magma. However, we note that these reactions do not encompass all the prominent H-C-N-O molecules at the considered conditions. In particular, NH$_3$ is expected to be the dominant N-bearing species at the base of the atmosphere. 
By not including NH$_3$ and its equilibrium with N$_2$ and H$_2$, S24 may be overestimating the nitrogen depletion in the atmosphere, given that all of the nitrogen is assumed to be in N$_2$, which is very soluble in magma at reducing conditions, as we show in Figure~\ref{fig:solubilities} in Appendix~\ref{app:solubility}.

In our framework, described in Section~\ref{sec:methods/melt_interface}, we find that nitrogen depletion in the atmosphere increases by several orders of magnitude by not including NH$_3$. Ultimately, this highlights the importance of the completeness of the reactions and solubilities considered. Finally, we note that it is also possible to not have significant N depletion even in the presence of a molten surface depending on the pressure and temperature, as shown in Table~\ref{tab:extreme}.

\subsubsection{Atmospheric composition and observables}
The properties at the surface determine the composition in the upper layers of the atmosphere, and hence its observable characteristics. These are strongly influenced by model assumptions on elemental abundances. S24 allow the \ch{C}/\ch{H} ratio to vary between $0.01 \times$ solar and $100 \times$ solar, while keeping the \ch{N}/\ch{H} ratio fixed to solar, i.e., \ch{N}/\ch{H} $ = 6.76 \times 10^{-5}$ by number. This itself limits the \ch{NH3} log-mixing ratio to at most $\log{X_{\ch{NH3}}} \sim -4$, close to the upper bound of $-4.46$ found by \citet{madhusudhan_carbon-bearing_2023}, and biases the model by construction to allow for up to $100 \times$ more (or down to $100 \times$ fewer) \ch{C}-based molecules than \ch{N}-based ones. The dependence of the S24 model outcomes on the choice of \ch{C}/\ch{H} values is not reported. It should be noted that a $100\times$ enhancement or depletion of \ch{C}/\ch{H} without any change in \ch{N} may be difficult to reconcile with potential formation mechanisms.

We note two further points regarding the abundance of \ch{C}- and \ch{N}-bearing species predicted by S24. Firstly, S24 appear to indicate that the total abundances of carbon in their models reach up to $3.8$ wt$\%$ of the planet mass. It is, however, unclear how this may be compatible with their assumptions of a C/H ratio of at most $100 \times$ solar  and an \ch{H} mass fraction $\leq 1 \%$, given they adopt the \citet{asplund_chemical_2009} value for (C/H)$_\odot$, i.e., $3.2 \times 10^{-3}$ by mass. Secondly, as argued in Section~\ref{subsubsec:interior}, only the largest atmospheric mass fractions S24 consider can potentially lead to a magma ocean. At the resulting surface pressures, however, their model predicts a log-mixing ratio for \ch{CO2} of $\mathrm{log} X_{\ch{CO2}} \lesssim -3$. This is at the lowest end, if not outside, of the $1 \sigma$ confidence interval presented in \citet{madhusudhan_carbon-bearing_2023}. Furthermore, the CO abundance or the \ch{CO2}/CO ratio are not reported in S24, making it difficult to assess the validity of the chemical estimates. 

Finally, S24 argue that the model spectra from their model ensemble provide a qualitatively reasonable match to the data. Even if the model spectra were taken at face value, the lack of a reported goodness-of-fit metric precludes a reliable assessment of the match to data. More generally, a limited grid of forward models is insufficient to robustly explore the full model space taking into account all the degeneracies involved in an atmospheric spectral model and to obtain a statistically robust fit to the data; that is the purpose of atmospheric retrievals \citep{madhusudhan_review_2018}. A more reliable approach in the present context is to compare the model-predicted chemical abundances with the abundance constraints obtained from robust atmospheric retrievals of the observed spectra. As discussed above, the cases of S24 with the highest surface pressure, i.e. those that may allow a magma surface, still predict lower CO$_2$ abundances than those retrieved for K2-18~b \citep{madhusudhan_carbon-bearing_2023}. The CO and H$_2$O abundances are not reported in S24, which prevents a clear assessment of the agreement between the chemical predictions and the retrieved abundances. 

\section{Results: A Case Study of K2-18~\lowercase{b}} \label{sec:results}

After having established the consistency of our results with \citet{gaillard_redox_2022}, and having discussed the S24 findings for K2-18~b, we proceed to apply our framework \textit{ex novo}. We do so for K2-18~b in the present section, starting, as outlined in Figure~\ref{fig:flowchart}, with internal and atmospheric structure modelling that ensures consistency with the known bulk parameters. 
Through considering magma-atmosphere interactions, equilibrium chemistry in the lower atmosphere and non-equilibrium processes in the upper atmosphere, we make predictions for the observable composition and spectral signatures of a sub-Neptune magma world.

\subsection{Atmospheric Structure}
As discussed in Section~\ref{sec:methods/atmospheric_structure}, the dayside atmospheric structure is calculated self-consistently from the atmospheric constraints retrieved in the one-offset case of \citet{madhusudhan_carbon-bearing_2023}: the median  $\log{X_{\ch{CH4}}} = -1.74$, $\log{X_{\ch{CO2}}} = -2.04$, and the 2$\sigma$ upper bound $\log{X_{\ch{H2O}}} = -3.01$. The $P$-$T$ profile depends on a wide range of parameters, not all of which are observationally well-constrained: these include the internal temperature $T_{\mathrm{int}}$, the properties of clouds/hazes if present, and the efficiency of day-night heat redistribution. A detailed exploration of the temperature profiles in deep \ch{H2}-rich sub-Neptune atmospheres has been carried out before, in \citet{piette_temperature_2020}. Here, we assume uniform day-night heat redistribution, and consider four cases for the $P$-$T$ profiles, varying the internal temperature $T_{\mathrm{int}}$ and the Rayleigh enhancement factor ($a$) for the hazes: C1, corresponding to $T_{\mathrm{int}} = 25$ K, $a = 10000$; C2 and C3, both with $a = 1500$, with $T_{\mathrm{int}} = 25$ K and $T_{\mathrm{int}} = 50$ K, respectively; C4, with $T_{\mathrm{int}} = 50$ K and $a = 100$. We note that, in principle, even colder profiles are plausible, given the clouds/haze properties retrieved from observations \citep{madhusudhan_carbon-bearing_2023}. All the $P$-$T$ profiles are shown in Figure~\ref{fig:interiorresults}.

\subsection{Internal Structure}

\begin{figure}
    \centering
    \includegraphics[width=\columnwidth]{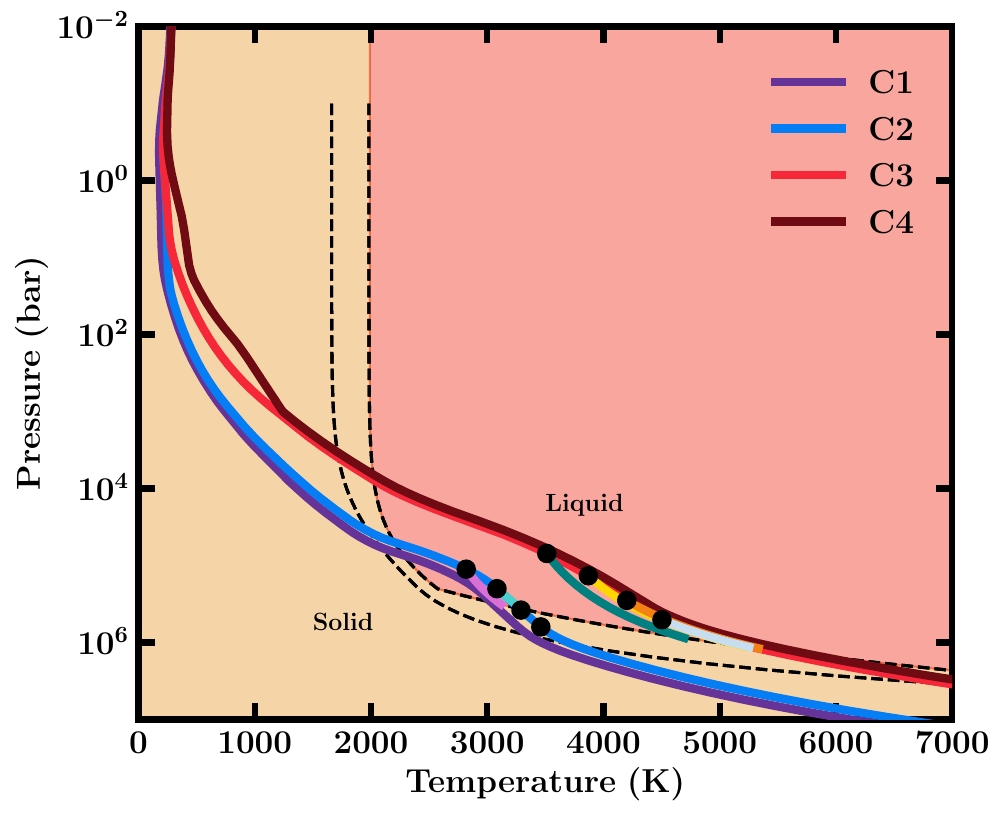}
    \caption{Atmospheric pressure-temperature profiles shown against the adopted phase boundary for the silicate mantle, with the dashed lines corresponding to the \citet{Monteux2016} liquidus and solidus for peridotite. The black circles indicate the surface conditions for the cases discussed and the coloured lines show the adiabatic temperature structure in the melt, adopting the liquidus as the melt-solid transition.}
    \label{fig:interiorresults}
\end{figure}

For each of these profiles, we obtain the permitted H$_2$-rich envelope mass fraction ($x_\mathrm{env}$) and corresponding surface conditions ($P_\mathrm{s}$, $T_\mathrm{s}$) based on the bulk properties of the planet, as discussed in Sections~\ref{sec:methods/internal_structure} and~\ref{subsec:bulkparameters} and shown in Table~\ref{tab:extreme}. We vary the interior composition from $f_\mathrm{silicate}=5\%$ to $f_\mathrm{silicate}=100\%$, adopting the median $M_\mathrm{p}=8.63\ \mathrm{M_\oplus}$ \citep{cloutier_confirmation_2019} and $R_\mathrm{p}=2.61\ \mathrm{R_\oplus}$ \citep{benneke_water_2019}. We note that the pure silicate and $95\%$ iron ($f_\mathrm{silicate}=5\%$) interior cases are unrealistic end-member interior compositions, but we consider them for completeness. We adopt $P_0=0.05$~bar as the outer boundary condition for the internal structure modelling, corresponding to the pressure at $R_\mathrm{p}$, based on \citet{madhusudhan_interior_2020}.

In Figure~\ref{fig:interiorresults} we show the $P$-$T$ profiles considered, along with the surface conditions (black circles) and adiabatic profiles in the melt for our nominal C2 and C3 scenarios, which we further discuss below. The results for all $P$-$T$ profiles are given in Table~\ref{tab:extreme}.

The presence and amount of magma depend on the adopted $P$-$T$ profile. We start by considering one of the colder profiles, C2. For an Earth-like interior, we find $x_\mathrm{env}=3.76\%$, with surface conditions $P_\mathrm{s}=2.00\times10^5$~bar and $T_\mathrm{s}=3084$~K. The melt mass fraction ($x_\mathrm{melt}$) in this case is $1.81\%$. For a Mercury-like interior, i.e. with higher Fe content, we find $x_\mathrm{env}=5.29\%$, with surface conditions $P_\mathrm{s}=3.79\times10^5$~bar and $T_\mathrm{s}=3290$~K. Based on our assumption of the liquidus as the melt curve, we class this as having $0\%$ melt in Table~\ref{tab:extreme}. In reality, these surface conditions lie between the liquidus and solidus, which would lead to a partially molten surface. This is also the case for the $f_\mathrm{silicate}=5\%$ interior, with $x_\mathrm{env}=6.60\%$, with surface conditions $P_\mathrm{s}=6.27\times10^5$~bar and $T_\mathrm{s}=3461$~K. On the other hand, for the extreme case of a pure silicate interior, we find a melt mass fraction of $3.16\%$, for $x_\mathrm{env}=2.62\%$, $P_\mathrm{s}=1.12\times10^5$~bar and $T_\mathrm{s}=2819$~K.

We next consider the higher-temperature $P$-$T$ profile C3, which permits solutions with a magma ocean surface for all the interior compositions considered. For each interior composition, the permitted envelope mass fraction, and hence the surface pressure, is lower for this hotter $P$-$T$ profile. For an Earth-like interior, we find $x_\mathrm{env}=2.52\%$, with surface conditions $P_\mathrm{s}=1.36\times10^5$~bar and $T_\mathrm{s}=3870$~K. The melt mass fraction in this case is $10.16\%$. For a Mercury-like interior, we find $x_\mathrm{env}=3.83\%$, with surface conditions $P_\mathrm{s}=2.83\times10^5$~bar and $T_\mathrm{s}=4200$~K. The corresponding $x_\mathrm{melt}$ is $5.78\%$. For the extreme case of a pure silicate interior, we find a lower $x_\mathrm{env}=1.62\%$, with $P_\mathrm{s}=6.97\times10^4$~bar and $T_\mathrm{s}=3512$~K. The melt mass fraction in this case is larger, at $11.91\%$. For the other extreme of $f_\mathrm{silicate}=5\%$, we obtain $x_\mathrm{env}=5.06\%$, for $P_\mathrm{s}=5.05\times10^5$~bar and $T_\mathrm{s}=4503$~K, with $x_\mathrm{melt}=2.62\%$. We note that including modelling of partial melt would somewhat increase the melt mass fraction in all cases.

As shown in Table 1, the envelope mass fractions and surface conditions we find for profiles C1 and C4 are very similar to C2 and C3 respectively. This is despite the differences in envelope temperature structure resulting from differing haze properties; the difference between C2 and C3 is primarily due to the differing $T_\mathrm{int}$.

\begin{figure*}
	\includegraphics[width=0.5\textwidth]{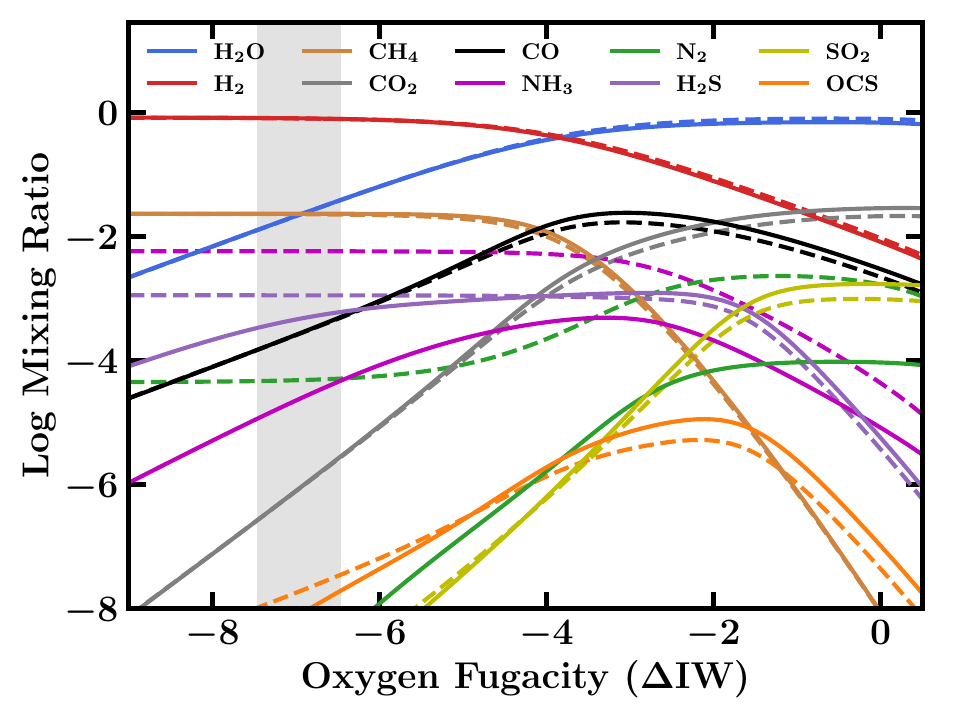}
    \includegraphics[width=0.5\textwidth]{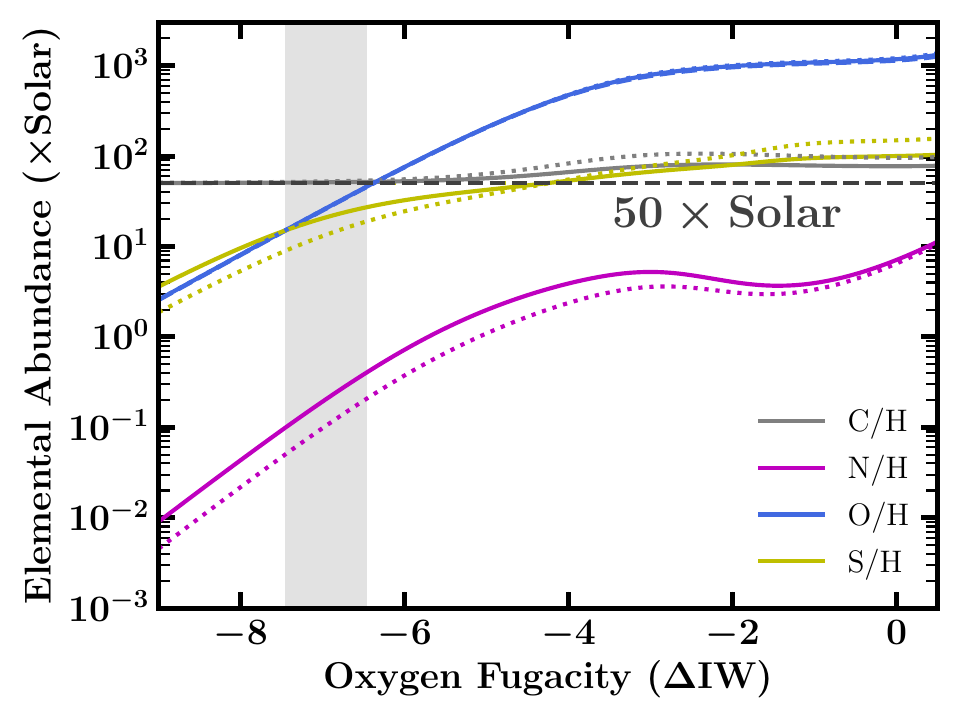}
    \caption{Atmospheric composition at the melt-atmosphere interface as a function of oxygen fugacity, at $T_\mathrm{s} = 4200$~K, $P_\mathrm{s}=2.83\times10^5$~bar, $M_\mathrm{melt} = 0.0578\,M_\mathrm{p}$, $M_\mathrm{atm} = 0.0383\,M_\mathrm{p}$ and 50$\times$solar metallicity -- corresponding to the C3 profile with a Mercury-like interior composition ($f_\mathrm{silicate}=30 \%$) in Table~\ref{tab:extreme}. {\bf Left:} Atmospheric mixing ratios of major H-C-O-N-S species. The solid and dashed lines show the abundances with and without solubility, respectively. {\bf Right:} Atmospheric elemental abundances normalised to hydrogen. The dotted line corresponds to a case with twice the melt mass fraction to highlight the potential effect of partial melt. The dashed horizontal line shows 50$\times$solar. The grey region centred at IW-7 ($\pm 0.5$), shown in both figures, corresponds to the approximate oxygen fugacity obtained with a total oxygen budget of 50$\times$solar.}
    \label{fig:melt} 
\end{figure*}

\subsection{Volatile Abundances at the Interface}
\label{sec:results/interface}

At the surface-atmosphere interface, the interactions between the gas phase equilibrium reactions and solubility of the gases in the magma, if any is present, drive the elemental abundances in the atmosphere. We consider the four $P$-$T$ profiles presented in Table~\ref{tab:extreme} and assume 50$\times$solar metallicity, using solar abundances by \cite{Asplund2021}. The assumed metallicity is approximately based on the median retrieved \ch{CH4} abundance for K2-18~b \citep{madhusudhan_carbon-bearing_2023}. Across all considered cases, we find that the dominant H-C-N-O-S gas species at the surface are $\ch{H2}$, $\ch{H2O}$, $\ch{CH4}$, $\ch{NH3}$, and $\ch{H2S}$. The resulting atmospheric elemental abundances from these scenarios are shown in Table~\ref{tab:extreme}. As expected, the atmosphere is highly reduced, with oxygen fugacities varying between IW-8.8 and IW-4.9 \citep[using the oxygen fugacity of the IW buffer by][]{Hirschmann2021} among the 12 cases with magma. We note that although our calculations of the oxygen fugacities agree to within 0.35 dex with the self-consistent IVTANTHERMO code at $P_\mathrm{s}=10^4$~bar and $T_\mathrm{s}=3000$~K, as described in Section~\ref{sec:methods/melt_interface}, the redox state at higher pressures/temperatures is not well understood. Future work is needed to better understand the redox state of gas dwarfs at these conditions. 

Overall, we find that \ch{H2O} and molecules containing \ch{N} and \ch{S} are the most dominant volatile species in the magma ocean, with high surface pressures strongly favouring the solubility of \ch{N2}. As such, for a given interior composition, we find that cooler $P$-$T$ profiles, resulting in higher $P_\mathrm{s}$, act to increase the depletion of nitrogen in the atmosphere - until the temperature is too low to support a molten surface. The dependence of N depletion on $P_\mathrm{s}$ is stronger than that on the melt fraction. Therefore, a hotter temperature profile does not necessarily result in higher N depletion. In terms of the internal structure, we find that the interior needs to be more iron-rich than Earth's interior to result in nitrogen depletion larger than $\sim$2 dex.

\begin{figure*}
    \centering
    \includegraphics[width=1.0\textwidth]{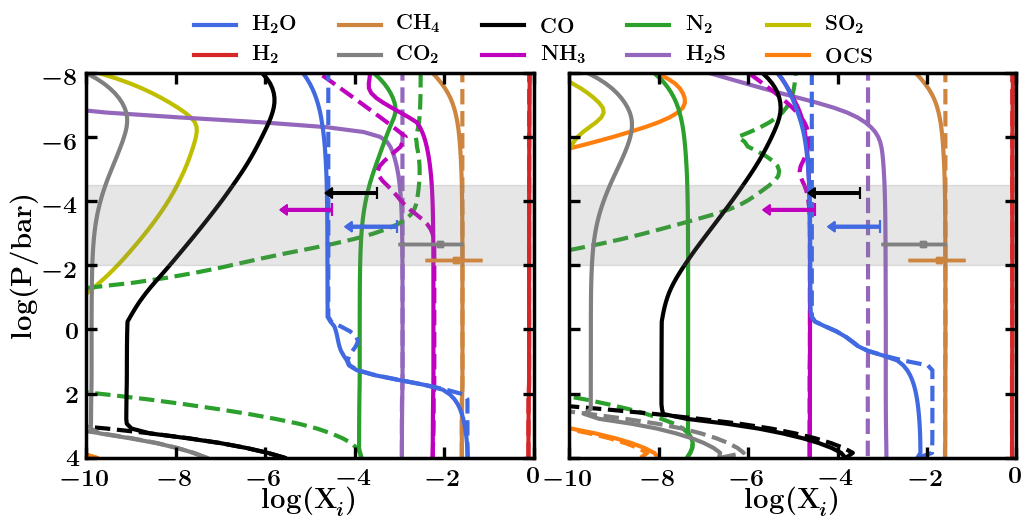}
    \caption{Vertical mixing ratio profiles for several H-C-O-N-S molecular species. ${\rm X}_i$ denotes the volume mixing ratio of a species $i$. Solid and dashed lines denote profiles computed with and without disequilibrium effects. Horizontal data points and arrows denote the mixing ratio constraints and 95\% upper estimates retrieved by \citet{madhusudhan_carbon-bearing_2023}. The gray shaded region denotes the pressure range typically probed by transmission spectroscopy \citep[e.g.][]{Constantinou2024}. {\bf Left:} Mixing ratio profiles corresponding to the C2 $P$-$T$ profile and 30\% interior silicate fraction case shown in Table \ref{tab:extreme}. This corresponds to a 50$\times$solar elemental abundances. {\bf Right:}  Profiles computed for the C3 $P$-$T$ profile and 30\% interior silicate fraction case shown in Table \ref{tab:extreme}. N is depleted due to dissolution in the magma surface.}
    \label{fig:Xprofiles}
\end{figure*}

Whilst we find that nitrogen can be depleted under certain conditions, in line with previous works investigating the solubility of nitrogen in reduced interiors \citep{daviau_experimental_2021, dasgupta_fate_2022, suer_distribution_2023, shorttle_distinguishing_2024}, we do not reproduce the six orders of magnitude depletion found by S24. Additionally, we also identify sulfur as a potential atmospheric tracer of a magma ocean; however, the depletion is less than that of nitrogen. Finally, we find that the solubility of H$_2$, CO, CO$_2$, and CH$_4$ is less prominent at the considered conditions and does not drive the abundances of these species far from chemical equilibrium expectations without a magma ocean. However, we note that, as further detailed in Appendix~\ref{app:solubility}, many molecular species lack solubility data at the extreme conditions considered here. Hence, further work is needed to improve our knowledge of the solubility of prominent volatiles in silicate melts.

In Figure~\ref{fig:melt}, we show the mixing ratios of the major C-H-O-N-S species in the lower atmosphere and the corresponding elemental abundances for a range of oxygen fugacities using the C3 profile and a Mercury-like interior ($f_\mathrm{silicate}=30 \%$). This represents the case with the strongest nitrogen depletion, excluding the extreme 5\% silicate interior cases, with atmospheric N/H being $\sim$2.5 dex lower than the assumed metallicity of 50$\times$solar. We also see the onset of sulfur depletion in the atmosphere due to the solubility of \ch{S2} at very reducing conditions ($\sim$IW-6 in this case). On the other hand, the carbon abundance remains unchanged, as mentioned above. We also highlight the potential effect of partial melt by doubling the melt mass fraction, shown by the dotted line in Figure~\ref{fig:melt}, leading to an approximately linear increase in the depletion of nitrogen.

\subsection{Atmospheric Chemistry}
\label{subsec:atmospheric_chemistry_calculations}

We now use the elemental abundances obtained above to determine the atmospheric composition above the surface, using equilibrium and non-equilibrium calculations. From across all the models shown in Table~\ref{tab:extreme}, we focus on two realistic cases, one with and one without melt. For the molten case, we consider the C3 profile with 30\% silicate fraction, which gives a significant N depletion. For the case with no melt we consider the C2 profile with 30\% silicate which has no N depletion. For each case, we set the atmospheric elemental budget to that obtained in Section~\ref{sec:results/interface} and reported in Table~\ref{tab:extreme}. As expected from the model set-up, the no-melt scenario results in all elemental abundances being identical to those of a 50$\times$solar metallicity gas. 

Across all cases considered, the primary O, C, N, and S reservoirs are H$_2$O, CH$_4$, NH$_3$ and H$_2$S over most of the atmosphere, as indicated by the dashed lines in Figure~\ref{fig:Xprofiles}. This is seen for a pressure range spanning over 10 orders of magnitude and a temperature profile ranging between $\sim$260-2700~K. 

\begin{figure*}
    \centering
    \includegraphics[width=\textwidth]{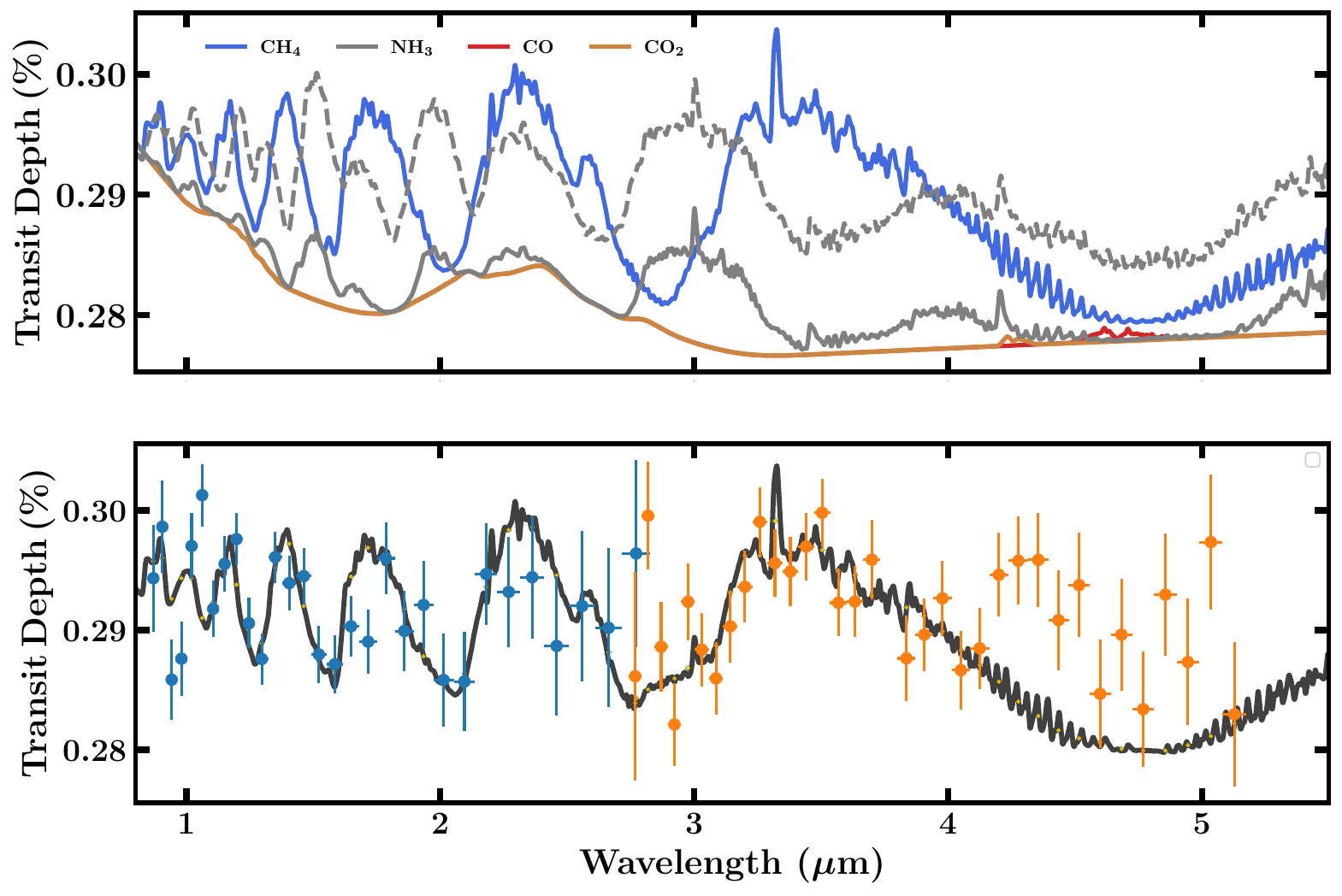}
    \caption{{\bf Top:} Spectral contributions arising from CH$_4$, NH$_3$, CO and CO$_2$ in the transmission spectrum of K2-18~b. The atmospheric abundance of each molecule corresponds to the right-hand side plot of Figure~\ref{fig:Xprofiles}, generated with the C3 $P$-$T$ profile, and includes disequilibrium effects. The dashed grey line corresponds to the spectral contribution of NH$_3$ if it were not depleted by dissolution in the magma. {\bf Bottom:} The resulting transmission spectrum from all four species' spectral contributions. Blue and orange errorbars are JWST NIRISS and NIRSpec G395H observations of K2-18~b, which include the -41 ppm retrieved offset reported  by \citet{madhusudhan_carbon-bearing_2023}. It can be seen that the magma ocean scenario does not result in sufficient CO$_2$ to explain the observations at $\sim$4.3~$\mu$m. We emphasise that the present comparison to the data is solely for illustration. A robust comparison necessitates considering the constraints obtained from a detailed retrieval analysis, as done in Figure \ref{fig:Xprofiles}.}
    \label{fig:spectral_contributions}
    \vspace{2mm}
\end{figure*}

The mixing ratio profiles obtained for the no-melt case are shown on the left-hand-side of Figure \ref{fig:Xprofiles}. In both the equilibrium and disequilibrium cases, the abundance of H$_2$O in the upper atmosphere is significantly depleted by a cold trap below the $\sim$1~bar pressure level. While CO and CO$_2$ are absent from the photosphere in the equilibrium case, they are present in the disequilibrium case, arising from photochemical processes. However, their abundance is significantly hindered by the limited availability of O, with the main carrier H$_2$O being depleted by the cold trap. The abundance of CO$_2$ is lower than that of CO throughout the atmosphere.

Compared to the retrieved atmospheric composition of K2-18~b \citep{madhusudhan_carbon-bearing_2023}, shown as errorbars and arrows in Figure \ref{fig:Xprofiles}, the computed CH$_4$ abundance is consistent with the retrieved constraint. However, there is a substantial difference of $\sim$8~dex between the computed abundance of CO$_2$ with the measured value across the observable pressure range. Additionally, the retrieved upper limits for H$_2$O and CO are consistent with the computed amounts. Lastly, the computed value of NH$_3$ is higher than, and therefore inconsistent with the retrieved upper limit.

The right-hand-side of Figure \ref{fig:Xprofiles} shows the case with a molten surface. This configuration results in very similar abundances for O- and C-carrying molecules as the no-melt case. This includes the significant depletion of H$_2$O due to a cold trap, the limited production of CO and CO$_2$, and CO being more abundant than CO$_2$.  The main difference from the no-melt case is the notable depletion of NH$_3$, due to N dissolving in the magma. Specifically NH$_3$ and N$_2$ are at much lower mixing ratios than in the no-melt case, by $\sim$2~dex. Compared to constraints from observations, CH$_4$, H$_2$O, CO and in this case NH$_3$ as well are consistent with the retrieved constraints and upper limits. However, the resulting CO$_2$ abundance is still substantially lower than the observed abundance.

In summary, we find that even for the case with significant melt, corresponding to our hotter $P$-$T$ profile with a high $T_{\rm int}$, the NH$_3$ abundance is close to the observed 95\% upper limit, while the CO$_2$ abundance is still significantly discrepant from the observed value and lower than CO. Therefore, we find that the retrieved atmospheric composition of K2-18~b \citep{madhusudhan_carbon-bearing_2023} is inconsistent with a magma ocean scenario, or more generally with a deep H$_2$-rich atmosphere with or without melt, for this planet. In principle, the absence of a cold trap could lead to higher H$_2$O abundance in the troposphere, which in turn could lead to higher CO$_2$ abundance. However, such a scenario would also give rise to a significant amount of H$_2$O and CO, which are presently not detected.

\subsection{Sensitivity to Atmospheric Parameters} \label{sec:results/5}

We also explore other values for the three key atmospheric parameters that may influence the observable composition: the metallicity, the eddy diffusion coefficient $K_\mathrm{zz}$, and the internal temperature $T_\mathrm{int}$. We consider the two cases shown in Figure~\ref{fig:Xprofiles} as the canonical cases corresponding to the two $P$-$T$ profiles (C2 and C3). Both cases assume a median metallicity of 50$\times$solar and $K_\mathrm{zz}$ of 10$^6$ cm$^2$s$^{-1}$ in the deep atmosphere. It may be argued that a higher metallicity could result in higher CO$_2$ abundances than the canonical cases and better match the observed abundances. Similarly, a broader range of $K_\mathrm{zz}$ may also influence the abundances. Therefore, for each of the two canonical cases, we investigate models with different values for the metallicity and $K_\mathrm{zz}$. We consider metallicities of 100$\times$solar and 300$\times$solar, representing cases with significantly higher metallicities beyond the median retrieved value of $\sim$50$\times$solar. For $K_\mathrm{zz}$, we explore two end-member scenarios of 10$^4$ cm$^2$s$^{-1}$ and 10$^8$ cm$^2$s$^{-1}$. Based on \citet{madhusudhan_interior_2020} and \citet{Valencia2013}, for our canonical cases we considered values of $25$~K and $50$~K for $T_\mathrm{int}$. We additionally consider the effect of using a $P$-$T$ profile with a higher $T_\mathrm{int}$ of $60$~K, as has been considered by \citet{Hu2021a}. As found for our canonical cases, we find that the observed CO and CO$_2$ abundances remain unexplained by these models with different values of metallicity, $K_\mathrm{zz}$ and $T_\mathrm{int}$. These results are discussed in full in Appendix~\ref{app:atmosphericchem}.

\subsection{Spectral Characteristics}

We use the atmospheric compositions computed in Section~\ref{subsec:atmospheric_chemistry_calculations} to examine the spectral signatures of CH$_4$, NH$_3$, CO and CO$_2$, which have been previously identified as key diagnostics of the presence of a magma surface. Using the VIRA retrieval framework's \citep{Constantinou2024} capability of considering non-uniform vertical mixing ratios, we directly use the atmospheric composition profiles computed using the VULCAN \citep{Tsai2021} non-equilibrium code described above and shown in Figure~\ref{fig:Xprofiles}. We specifically consider the melt case discussed above, to evaluate the spectral implications for the presence of a magma layer. For all cases, we consider parametric grey cloud and Rayleigh-like haze properties corresponding to the median retrieved constraints of \citet{madhusudhan_carbon-bearing_2023}, to facilitate a qualitative comparison with the observations. Specifically, we set log($a$) = $10^{7.31}$, $\gamma = -11.67$, P$_\mathrm{c} = 10^{-0.55}$ and $\phi_\mathrm{c} = 0.63$.

The resulting spectral contributions and transmission spectrum are shown in Figure~\ref{fig:spectral_contributions}. As can be seen in the top panel, CH$_4$ has prominent spectral features throughout the 1-5 $\mu$m wavelength range, while CO$_2$ and CO give rise to absorption features at $\sim$4.3 and $\sim$4.7~$\mu$m respectively. NH$_3$ shows a spectral feature at $\sim$3~$\mu$m. Due to the depletion of atmospheric nitrogen arising from its dissolution in the magma, the NH$_3$ spectral contribution is relatively weak and not detected in the present data. Without such a depletion, i.e. with nitrogen at a solar elemental abundance ratio, NH$_3$ would have prominent spectral features across the wavelength range of comparable strength to CH$_4$. While CO is more abundant than CO$_2$ in the observable atmosphere, as described in Section~\ref{subsec:atmospheric_chemistry_calculations}, the low absolute abundances of both molecules give rise to comparably weak spectral features.

The resulting transmission spectrum provides a reasonable match for the NIRISS observations of K2-18 b at shorter wavelengths due to the strong \ch{CH4} features. However, the spectrum does not fit the prominent CO$_2$ absorption feature seen in the NIRSpec G395H data. Moreover, the spectral contribution of CO is also minimal. Together, the two molecules are present at abundances that do not provide a good fit to the data in the 4-5 $\mu$m range. 

Overall, we find that the gas dwarf scenario of a thick H$_2$-rich atmosphere of K2-18 b in equilibrium with a magma ocean at depth is not consistent with the existing JWST observations. In particular, irrespective of the NH$_3$ depletion, the models predict a low CO$_2$ abundance and CO $>$ CO$_2$ which are inconsistent with the retrieved abundances. Future studies need to investigate if other effects may contribute to the observed composition. For example, similar to that  discussed in \citet{madhusudhan_carbon-bearing_2023}, in order for the detected abundance of \ch{CO2} to be compatible with a deep H$_2$-rich atmosphere, an unphysically low C/O ratio of $\sim$ 0.02--0.06, together with a moderate C/H ratio ($\sim$30-50$\times$ solar) and vertical quenching may be required. However, such an atmosphere could also lead to significant CO abundances that may not be consistent with the observations, and the deep atmosphere would have more H$_2$O than H$_2$.

\section{Summary and Discussion} \label{sec:discussion}

In this study, we report an integrated framework to investigate the plausibility of magma oceans on temperate gas dwarfs, and their potential atmospheric  signatures. Our framework models the various components of a planet, and their interplay. Specifically, it includes atmospheric and internal structure modelling, magma-atmosphere chemical interactions, and equilibrium as well as disequilibrium (photochemistry and vertical mixing) processes in the atmosphere. Considering all these coupled factors, it predicts the observable abundances of molecular species in the atmosphere and the expected spectral features. 

We apply our framework to perform a comparative assessment of previous works, validating our modelling of magma-atmosphere interactions against \citet{gaillard_redox_2022} and assessing the model predictions of \citet{shorttle_distinguishing_2024}, S24, for a temperate sub-Neptune. Our findings highlight the importance of considering physically plausible models, set up in a holistic framework. In particular, we note that the use of stand-alone magma-atmosphere interaction models, which do not consider the complex interplay of interior and atmospheric factors, can lead to erroneous results. 

\subsection{Summary}

Magma oceans are normally expected for rocky planets with high equilibrium temperatures. In the present work, we have tested the limits of this scenario by exploring whether K2-18~b, a habitable-zone sub-Neptune, can host a magma ocean, as previously suggested by S24, and what the observable signatures could be. We summarise our key findings as follows: 

\begin{itemize}
    \item An integrated framework is essential to obtain physically plausible and self-consistent results for modelling sub-Neptune gas dwarfs. Our framework includes an atmosphere and interior structure model, including phase diagrams and equations of state of appropriate silicates; thermochemical equilibrium calculations for the silicates-atmosphere interface and lower atmosphere; and disequilibrium processes throughout the atmosphere.
    
    \item The melt fraction admissible in a gas dwarf depends on atmospheric and interior properties, specifically the interior composition and the atmospheric $P$-$T$ profile. The $P$-$T$ profile, in turn, depends strongly on the internal temperature $T_{\mathrm{int}}$, as well as on the presence and properties of clouds/hazes and on the molecular absorbers present in the atmosphere. For a gas dwarf scenario assuming the bulk parameters of K2-18~b, we find that, with an Earth-like interior composition, maximal melt mass fractions of $\sim$10\% are possible, and may increase somewhat if partial melting is considered. 

    \item A planet's bulk parameters and temperature structure place both upper and lower limits on the envelope mass fraction, assuming a gas dwarf scenario. For the K2-18~b models considered in this work, these limits are $\sim$1\% and $7$\% of the planet mass, corresponding to a pure silicate and a $95\%$ iron interior, respectively. The envelope mass fraction affects the surface pressure at the rock-atmosphere boundary, which, in turn, affects the potential melt conditions. 
    
    \item We find using our framework that the current chemical constraints for K2-18~b are inconsistent with a magma ocean scenario or any gas dwarf scenario, contrary to S24. Firstly, the high observed abundance of \ch{CO2} along with low \ch{H2O} is inconsistent with the chemical expectations for the gas dwarf scenario. Secondly, we find CO to be higher than \ch{CO2} by over 1 dex which is also inconsistent with the observations. We find this to be the case with or without a magma ocean, and relatively  independent of the uncertainties in magma-atmosphere interactions at the extremely reduced conditions as described in Appendix \ref{app:solubility}. Finally, we find that N depletion in the atmosphere depends on a wide range of atmospheric and interior parameters, and can range between no depletion and $\sim$2.5 dex for a realistic model space, given available solubility data. 
    
    \item Overall, we find that key atmospheric signatures for identifying a gas dwarf include the \ch{CO} and \ch{CO2} abundances, and, if melt is present, possible nitrogen depletion, consistent with some previous studies (cf. Section \ref{sec:previous}). In particular, we expect that \ch{CO}/\ch{CO2}$> 1$ if no \ch{H2O} is observed (as a result, e.g., of condensation), or, in the presence of \ch{H2O}, \ch{CO}/\ch{CO2} $\lesssim 1$, due to photolysis of \ch{H2O} making more oxygen available for the formation of \ch{CO2}. Furthermore, we find that N depletion is more sensitive to the surface pressure than to the amount of melt present, provided this is non-zero. Thus, the presence of a magma ocean does not ensure a significant N depletion in the atmosphere.

    \item Our models predict significant \ch{H2S} for a deep \ch{H2}-rich atmosphere scenario. Hence, a lack of \ch{H2S} may be indicative of a shallow atmosphere. However, we note that there are significant uncertainties in the behaviour of S-bearing species in silicate melts at such extremely reducing conditions. Therefore, more robust data for such conditions is needed in order for this signature to be used with a higher degree of confidence. We also note that there is uncertainty in the sulfur photochemical network for such planetary conditions.

    \item As discussed below, a number of important unknowns remain. In particular, as discussed in Appendix~\ref{app:solubility}, the solubility of \ch{NH3} in magma remains poorly understood, especially at extremely reducing conditions, as is also the case for \ch{H2S} at high pressures and temperatures.

\end{itemize}

\subsection{Future Work}
\label{sec:conclusion/future}
In order to aid accurate modelling of potential gas dwarf magma ocean planets, further developments are needed in three areas: (1) solubility laws for volatiles at extremely high pressures and temperatures and very reducing conditions, (2) equations of state (EOS) of silicates at the conditions relevant to temperate sub-Neptunes, and (3) complete reactions lists for all relevant atmospheric species.

As discussed in Appendix~\ref{app:solubility}, there is a pressing need for further experimental data and/or ab initio simulations on the solubility of volatile species in silicate melt at the physical and chemical conditions that we have shown in this study to be relevant to the magma-atmosphere interface on sub-Neptunes. This includes high pressure and temperature, and low oxygen fugacity. In particular, the availability of \ch{NH3} solubility laws at these conditions would allow more precise prescriptions than assuming its solubility to be negligible, avoiding the resulting likely overestimation of the abundance of N-bearing species in the atmosphere. 
In general, present laws are expected to give an order-of-magnitude estimate of the solubility at the conditions explored in this study; future work is needed to improve the solubility data.

Furthermore, once more accurate and precise solubility laws become available, the non-ideality of gas behaviour at the high pressures relevant at the interface may become a notable source of error if ignored, and will thus need to be appropriately treated \citep{kite_superabundance_2019, schlichting_chemical_2022}. We also note that, as a result of the lack of knowledge on the solubility of volatiles in the melt, the phase of the melt itself is not well-constrained. In particular, it is possible that some of the models considered here fall in a regime where there is no surface, and the atmosphere and magma become a single continuous phase at some lower pressure. This would happen if the volatiles were completely miscible in the melt, as is the case for water above a few GPa \citep{ni_supercritical_2017}. It is however not known whether this behaviour applies to \ch{H2}-dominated atmospheres such as the one considered here. Furthermore, even if complete miscibility is not achieved, it is possible that the presence of volatiles in the magma may lead to a change in its EOS, which has not been accounted for here, where we have instead assumed a volatile-free melt for the internal structure calculations.  

There is also scope for future work on the internal structure modelling, including the melt. This includes implementing the partial melting that would occur due to the magma's heterogeneous nature between the solidus and liquidus, as shown in Figures~\ref{fig:surfaceconds} and~\ref{fig:interiorresults}. This is expected to result in a larger fraction of the mantle being at least partially melted than when considering the fully melted region alone, hence further depleting the atmosphere of the most soluble species. This effect is however in part addressed in this work, by considering the impact of a doubled melt mass fraction, as shown in Figure~\ref{fig:melt}. Furthermore, future work will include more detailed prescriptions for the mantle, including alternate mineral compositions, and a fully temperature-dependent EOS for the solid portion. 

Overall, JWST provides a promising avenue for atmospheric characterisation of sub-Neptune exoplanets. The high quality of the observations means that concomitant advances need to be made in theoretical models to maximise the scientific return from the data. In this work, we have outlined an end-to-end framework for gas dwarf sub-Neptunes to enable an evaluation of this scenario given high precision JWST data, and highlight the need for more accurate inputs for these models. Such advancements in both observations and theory promise a new era in the characterisation of low-mass exoplanets with JWST.

\FloatBarrier

%\section{Acknowledgements}
%\begin{acknowledgments}

\vspace{4mm}
\textit{Acknowledgements:} We thank the reviewer for their helpful comments on the manuscript. N.M., L.P.C. and F.R. acknowledge support from the Science \& Technologies Facilities Council (STFC) towards the PhD studies of L.P.C. and F.R. (UKRI grants 2886925 and 2605554). N.M. and M.H. acknowledge support from STFC Center for Doctoral Training (CDT) in Data Intensive Science at the University of Cambridge (grant No. ST/P006787/1), and the MERAC Foundation, Switzerland, towards the doctoral studies of M.H. N.M. and S.C. acknowledge support from the  UK Research and Innovation (UKRI) Frontier Grant (EP/X025179/1, PI: N. Madhusudhan). J.D. acknowledges support from Grant EAR‐ 2242946 of National Science Foundation. K.K.M.L acknowledges support from the US Coast Guard Academy Faculty Research Fellowship. J.M. acknowledges NASA grant 80NSSC23K0281. This work was performed using resources provided by the Cambridge Service for Data Driven Discovery (CSD3) operated by the University of Cambridge Research Computing Service (\url{www.csd3.cam.ac.uk}), provided by Dell EMC and Intel using Tier-2 funding from STFC (capital grant EP/P020259/1), and DiRAC funding from STFC (\url{www.dirac.ac.uk}).

%\end{acknowledgments}

\bibliography{ms}{}

\begin{thebibliography}{}
\expandafter\ifx\csname natexlab\endcsname\relax\def\natexlab#1{#1}\fi
\providecommand{\url}[1]{\href{#1}{#1}}
\providecommand{\dodoi}[1]{doi:~\href{http://doi.org/#1}{\nolinkurl{#1}}}
\providecommand{\doeprint}[1]{\href{http://ascl.net/#1}{\nolinkurl{http://ascl.net/#1}}}
\providecommand{\doarXiv}[1]{\href{https://arxiv.org/abs/#1}{\nolinkurl{https://arxiv.org/abs/#1}}}

\bibitem[{{Abel} {et~al.}(2011){Abel}, {Frommhold}, {Li}, \& {Hunt}}]{Abel2011}
{Abel}, M., {Frommhold}, L., {Li}, X., \& {Hunt}, K. L.~C. 2011, Journal of Physical Chemistry A, 115, 6805, \dodoi{10.1021/jp109441f}

\bibitem[{Ardia {et~al.}(2013)Ardia, Hirschmann, Withers, \& Stanley}]{ardia_solubility_2013}
Ardia, P., Hirschmann, M.~M., Withers, A.~C., \& Stanley, B.~D. 2013, Geochimica et Cosmochimica Acta, 114, 52, \dodoi{10.1016/j.gca.2013.03.028}

\bibitem[{Armstrong {et~al.}(2015)Armstrong, Hirschmann, Stanley, Falksen, \& Jacobsen}]{armstrong_speciation_2015}
Armstrong, L.~S., Hirschmann, M.~M., Stanley, B.~D., Falksen, E.~G., \& Jacobsen, S.~D. 2015, Geochimica et Cosmochimica Acta, 171, 283, \dodoi{10.1016/j.gca.2015.07.007}

\bibitem[{{Asplund} {et~al.}(2021){Asplund}, {Amarsi}, \& {Grevesse}}]{Asplund2021}
{Asplund}, M., {Amarsi}, A.~M., \& {Grevesse}, N. 2021, \aap, 653, A141, \dodoi{10.1051/0004-6361/202140445}

\bibitem[{Asplund {et~al.}(2009)Asplund, Grevesse, Sauval, \& Scott}]{asplund_chemical_2009}
Asplund, M., Grevesse, N., Sauval, A.~J., \& Scott, P. 2009, Annual Review of Astronomy and Astrophysics, 47, 481, \dodoi{10.1146/annurev.astro.46.060407.145222}

\bibitem[{{Azzam} {et~al.}(2016){Azzam}, {Tennyson}, {Yurchenko}, \& {Naumenko}}]{Azzam2016}
{Azzam}, A. A.~A., {Tennyson}, J., {Yurchenko}, S.~N., \& {Naumenko}, O.~V. 2016, \mnras, 460, 4063, \dodoi{10.1093/mnras/stw1133}

\bibitem[{{Barber} {et~al.}(2014){Barber}, {Strange}, {Hill}, {Polyansky}, {Mellau}, {Yurchenko}, \& {Tennyson}}]{Barber2014}
{Barber}, R.~J., {Strange}, J.~K., {Hill}, C., {et~al.} 2014, \mnras, 437, 1828, \dodoi{10.1093/mnras/stt2011}

\bibitem[{{Barber} {et~al.}(2006){Barber}, {Tennyson}, {Harris}, \& {Tolchenov}}]{Barber2006}
{Barber}, R.~J., {Tennyson}, J., {Harris}, G.~J., \& {Tolchenov}, R.~N. 2006, \mnras, 368, 1087, \dodoi{10.1111/j.1365-2966.2006.10184.x}

\bibitem[{Bean {et~al.}(2021)Bean, Raymond, \& Owen}]{bean_nature_2021}
Bean, J.~L., Raymond, S.~N., \& Owen, J.~E. 2021, Journal of Geophysical Research: Planets, 126, e2020JE006639, \dodoi{10.1029/2020JE006639}

\bibitem[{Belov {et~al.}(1999)Belov, Iorish, \& Yungman}]{BELOV1999173}
Belov, G.~V., Iorish, V.~S., \& Yungman, V.~S. 1999, Calphad, 23, 173, \dodoi{https://doi.org/10.1016/S0364-5916(99)00023-1}

\bibitem[{Benneke {et~al.}(2019)Benneke, Wong, Piaulet, Knutson, Lothringer, Morley, Crossfield, Gao, Greene, Dressing, Dragomir, Howard, McCullough, Kempton, Fortney, \& Fraine}]{benneke_water_2019}
Benneke, B., Wong, I., Piaulet, C., {et~al.} 2019, The Astrophysical Journal Letters, 887, L14, \dodoi{10.3847/2041-8213/ab59dc}

\bibitem[{{Benneke} {et~al.}(2024){Benneke}, {Roy}, {Coulombe}, {Radica}, {Piaulet}, {Ahrer}, {Pierrehumbert}, {Krissansen-Totton}, {Schlichting}, {Hu}, {Yang}, {Christie}, {Thorngren}, {Young}, {Pelletier}, {Knutson}, {Miguel}, {Evans-Soma}, {Dorn}, {Gagnebin}, {Fortney}, {Komacek}, {MacDonald}, {Raul}, {Cloutier}, {Acuna}, {Lafreni{\`e}re}, {Cadieux}, {Doyon}, {Welbanks}, \& {Allart}}]{Benneke2024}
{Benneke}, B., {Roy}, P.-A., {Coulombe}, L.-P., {et~al.} 2024, arXiv e-prints, arXiv:2403.03325, \dodoi{10.48550/arXiv.2403.03325}

\bibitem[{Bernadou {et~al.}(2021)Bernadou, Gaillard, Füri, Marrocchi, \& Slodczyk}]{bernadou_nitrogen_2021}
Bernadou, F., Gaillard, F., Füri, E., Marrocchi, Y., \& Slodczyk, A. 2021, Chemical Geology, 573, 120192, \dodoi{10.1016/j.chemgeo.2021.120192}

\bibitem[{{Borysow} {et~al.}(1988){Borysow}, {Frommhold}, \& {Birnbaum}}]{Borysow1988}
{Borysow}, J., {Frommhold}, L., \& {Birnbaum}, G. 1988, \apj, 326, 509, \dodoi{10.1086/166112}

\bibitem[{Boulliung {et~al.}(2020)Boulliung, Füri, Dalou, Tissandier, Zimmermann, \& Marrocchi}]{boulliung_oxygen_2020}
Boulliung, J., Füri, E., Dalou, C., {et~al.} 2020, Geochimica et Cosmochimica Acta, 284, 120, \dodoi{10.1016/j.gca.2020.06.020}

\bibitem[{Boulliung \& Wood(2023)}]{boulliung_sulfur_2023}
Boulliung, J., \& Wood, B.~J. 2023, Contributions to Mineralogy and Petrology, 178, 56, \dodoi{10.1007/s00410-023-02033-9}

\bibitem[{Bower {et~al.}(2022)Bower, Hakim, Sossi, \& Sanan}]{Bower2022}
Bower, D.~J., Hakim, K., Sossi, P.~A., \& Sanan, P. 2022, The Planetary Science Journal, 3, 93, \dodoi{10.3847/PSJ/ac5fb1}

\bibitem[{Castor {et~al.}(1992)Castor, Dykema, \& Klein}]{castor_new_1992}
Castor, J.~I., Dykema, P.~G., \& Klein, R.~I. 1992, ApJ. . ., 387

\bibitem[{{Chabrier} {et~al.}(2019){Chabrier}, {Mazevet}, \& {Soubiran}}]{Chabrier2019}
{Chabrier}, G., {Mazevet}, S., \& {Soubiran}, F. 2019, \apj, 872, 51, \dodoi{10.3847/1538-4357/aaf99f}

\bibitem[{Charnoz {et~al.}(2023)Charnoz, Falco, Tremblin, Sossi, Caracas, \& Lagage}]{charnoz_effect_2023}
Charnoz, S., Falco, A., Tremblin, P., {et~al.} 2023, Astronomy \& Astrophysics, 674, A224, \dodoi{10.1051/0004-6361/202245763}

\bibitem[{Chase(1998)}]{Chase1998}
Chase, M. 1998, NIST-JANAF Thermochemical Tables, 4th Edition (American Institute of Physics, -1)

\bibitem[{Chubb {et~al.}(2020)Chubb, Tennyson, \& Yurchenko}]{Chubb2020}
Chubb, K.~L., Tennyson, J., \& Yurchenko, S.~N. 2020, Monthly Notices of the Royal Astronomical Society, 493, 1531, \dodoi{10.1093/mnras/staa229}

\bibitem[{{Chubb} {et~al.}(2018){Chubb}, {Naumenko}, {Keely}, {Bartolotto}, {Macdonald}, {Mukhtar}, {Grachov}, {White}, {Coleman}, {Liu}, {Fazliev}, {Polovtseva}, {Horneman}, {Campargue}, {Furtenbacher}, {Cs{\'a}sz{\'a}r}, {Yurchenko}, \& {Tennyson}}]{Chubb2018}
{Chubb}, K.~L., {Naumenko}, O., {Keely}, S., {et~al.} 2018, \jqsrt, 218, 178, \dodoi{10.1016/j.jqsrt.2018.07.012}

\bibitem[{Clemente {et~al.}(2004)Clemente, Scaillet, \& Pichavant}]{clemente_solubility_2004}
Clemente, B., Scaillet, B., \& Pichavant, M. 2004, Journal of Petrology, 45, 2171, \dodoi{10.1093/petrology/egh052}

\bibitem[{{Cloutier} \& {Menou}(2020)}]{Cloutier2020}
{Cloutier}, R., \& {Menou}, K. 2020, \aj, 159, 211, \dodoi{10.3847/1538-3881/ab8237}

\bibitem[{Cloutier {et~al.}(2019)Cloutier, Astudillo-Defru, Doyon, Bonfils, Almenara, Bouchy, Delfosse, Forveille, Lovis, Mayor, Menou, Murgas, Pepe, Santos, Udry, \& Wünsche}]{cloutier_confirmation_2019}
Cloutier, R., Astudillo-Defru, N., Doyon, R., {et~al.} 2019, Astronomy \& Astrophysics, 621, A49, \dodoi{10.1051/0004-6361/201833995}

\bibitem[{{Constantinou} \& {Madhusudhan}(2024)}]{Constantinou2024}
{Constantinou}, S., \& {Madhusudhan}, N. 2024, \mnras, 530, 3252, \dodoi{10.1093/mnras/stae633}

\bibitem[{{Damiano} {et~al.}(2024){Damiano}, {Bello-Arufe}, {Yang}, \& {Hu}}]{Damiano2024}
{Damiano}, M., {Bello-Arufe}, A., {Yang}, J., \& {Hu}, R. 2024, \apjl, 968, L22, \dodoi{10.3847/2041-8213/ad5204}

\bibitem[{Dasgupta {et~al.}(2022)Dasgupta, Falksen, Pal, \& Sun}]{dasgupta_fate_2022}
Dasgupta, R., Falksen, E., Pal, A., \& Sun, C. 2022, Geochimica et Cosmochimica Acta, 336, 291, \dodoi{10.1016/j.gca.2022.09.012}

\bibitem[{Daviau \& Lee(2021)}]{daviau_experimental_2021}
Daviau, K., \& Lee, K. K.~M. 2021, Journal of Geophysical Research: Planets, 126, e2020JE006687, \dodoi{10.1029/2020JE006687}

\bibitem[{Dixon {et~al.}(1995)Dixon, Stolper, \& Holloway}]{dixon_experimental_1995}
Dixon, J.~E., Stolper, E.~M., \& Holloway, J.~R. 1995, Journal of Petrology, 36, 1607, \dodoi{10.1093/oxfordjournals.petrology.a037267}

\bibitem[{Dorn {et~al.}(2017)Dorn, Venturini, Khan, Heng, Alibert, Helled, Rivoldini, \& Benz}]{dorn_generalized_2017}
Dorn, C., Venturini, J., Khan, A., {et~al.} 2017, Astronomy \& Astrophysics, 597, A37, \dodoi{10.1051/0004-6361/201628708}

\bibitem[{Doyon(2024)}]{doyon_temperate_2024}
Doyon, R. 2024, Do {Temperate} {Rocky} {Planets} {Around} {M} {Dwarfs} have an {Atmosphere} ?,  arXiv, \dodoi{10.48550/arXiv.2403.12617}

\bibitem[{{Elkins-Tanton}(2008)}]{ElkinsTanton2008}
{Elkins-Tanton}, L.~T. 2008, Earth and Planetary Science Letters, 271, 181, \dodoi{10.1016/j.epsl.2008.03.062}

\bibitem[{Falco {et~al.}(2024)Falco, Tremblin, Charnoz, Ridgway, \& Lagage}]{falco_hydrogenated_2024}
Falco, A., Tremblin, P., Charnoz, S., Ridgway, J.~R., \& Lagage, P.-O. 2024, Astronomy \& Astrophysics, \dodoi{10.1051/0004-6361/202347650}

\bibitem[{{Fiquet} {et~al.}(2010){Fiquet}, {Auzende}, {Siebert}, {Corgne}, {Bureau}, {Ozawa}, \& {Garbarino}}]{Fiquet2010}
{Fiquet}, G., {Auzende}, A.~L., {Siebert}, J., {et~al.} 2010, Science, 329, 1516, \dodoi{10.1126/science.1192448}

\bibitem[{{Fortney} {et~al.}(2013){Fortney}, {Mordasini}, {Nettelmann}, {Kempton}, {Greene}, \& {Zahnle}}]{Fortney2013}
{Fortney}, J.~J., {Mordasini}, C., {Nettelmann}, N., {et~al.} 2013, \apj, 775, 80, \dodoi{10.1088/0004-637X/775/1/80}

\bibitem[{{Fressin} {et~al.}(2013){Fressin}, {Torres}, {Charbonneau}, {Bryson}, {Christiansen}, {Dressing}, {Jenkins}, {Walkowicz}, \& {Batalha}}]{Fressin2013}
{Fressin}, F., {Torres}, G., {Charbonneau}, D., {et~al.} 2013, \apj, 766, 81, \dodoi{10.1088/0004-637X/766/2/81}

\bibitem[{{Fulton} \& {Petigura}(2018)}]{Fulton2018}
{Fulton}, B.~J., \& {Petigura}, E.~A. 2018, \aj, 156, 264, \dodoi{10.3847/1538-3881/aae828}

\bibitem[{Fulton {et~al.}(2017)Fulton, Petigura, Howard, Isaacson, Marcy, Cargile, Hebb, Weiss, Johnson, Morton, Sinukoff, Crossfield, \& Hirsch}]{fulton_california-kepler_2017}
Fulton, B.~J., Petigura, E.~A., Howard, A.~W., {et~al.} 2017, The Astronomical Journal, 154, 109, \dodoi{10.3847/1538-3881/aa80eb}

\bibitem[{Gaillard {et~al.}(2022)Gaillard, Bernadou, Roskosz, Bouhifd, Marrocchi, Iacono-Marziano, Moreira, Scaillet, \& Rogerie}]{gaillard_redox_2022}
Gaillard, F., Bernadou, F., Roskosz, M., {et~al.} 2022, Earth and Planetary Science Letters, 577, 117255, \dodoi{10.1016/j.epsl.2021.117255}

\bibitem[{{Gandhi} \& {Madhusudhan}(2017)}]{Gandhi2017}
{Gandhi}, S., \& {Madhusudhan}, N. 2017, \mnras, 472, 2334, \dodoi{10.1093/mnras/stx1601}

\bibitem[{Gao {et~al.}(2022)Gao, Yang, Yang, \& Li}]{gao_experimental_2022}
Gao, Z., Yang, Y.-N., Yang, S.-Y., \& Li, Y. 2022, Geochimica et Cosmochimica Acta, 326, 17, \dodoi{10.1016/j.gca.2022.04.001}

\bibitem[{Ginzburg {et~al.}(2016)Ginzburg, Schlichting, \& Sari}]{ginzburg_super-earth_2016}
Ginzburg, S., Schlichting, H.~E., \& Sari, R. 2016, The Astrophysical Journal, 825, 29, \dodoi{10.3847/0004-637X/825/1/29}

\bibitem[{Ginzburg {et~al.}(2018)Ginzburg, Schlichting, \& Sari}]{ginzburg_core-powered_2018}
---. 2018, Monthly Notices of the Royal Astronomical Society, 476, 759, \dodoi{10.1093/mnras/sty290}

\bibitem[{{Glein}(2024)}]{glein_geochemical_2024}
{Glein}, C.~R. 2024, \apjl, 964, L19, \dodoi{10.3847/2041-8213/ad3079}

\bibitem[{Guillot \& Sator(2011)}]{guillot_carbon_2011}
Guillot, B., \& Sator, N. 2011, Geochimica et Cosmochimica Acta, 75, 1829, \dodoi{10.1016/j.gca.2011.01.004}

\bibitem[{Gupta \& Schlichting(2019)}]{gupta_sculpting_2019}
Gupta, A., \& Schlichting, H.~E. 2019, Monthly Notices of the Royal Astronomical Society, 487, 24, \dodoi{10.1093/mnras/stz1230}

\bibitem[{Gupta \& Schlichting(2020)}]{gupta_signatures_2020}
---. 2020, Monthly Notices of the Royal Astronomical Society, 493, 792, \dodoi{10.1093/mnras/staa315}

\bibitem[{Gurvich \& Veyts(1990)}]{gurvich1990thermodynamic}
Gurvich, L.~V., \& Veyts, I. 1990, Thermodynamic properties of individual substances: elements and compounds, Vol.~2 (CRC press)

\bibitem[{{Hamano} {et~al.}(2013){Hamano}, {Abe}, \& {Genda}}]{Hamano2013}
{Hamano}, K., {Abe}, Y., \& {Genda}, H. 2013, \nat, 497, 607, \dodoi{10.1038/nature12163}

\bibitem[{Hirschmann(2021)}]{Hirschmann2021}
Hirschmann, M. 2021, Geochimica et Cosmochimica Acta, 313, 74, \dodoi{https://doi.org/10.1016/j.gca.2021.08.039}

\bibitem[{Hirschmann {et~al.}(2012)Hirschmann, Withers, Ardia, \& Foley}]{hirschmann_solubility_2012}
Hirschmann, M., Withers, A., Ardia, P., \& Foley, N. 2012, Earth and Planetary Science Letters, 345-348, 38, \dodoi{10.1016/j.epsl.2012.06.031}

\bibitem[{Hirschmann(2016)}]{hirschmann_constraints_2016}
Hirschmann, M.~M. 2016, American Mineralogist, 101, 540, \dodoi{10.2138/am-2016-5452}

\bibitem[{{Holmberg} \& {Madhusudhan}(2024)}]{Holmberg2024}
{Holmberg}, M., \& {Madhusudhan}, N. 2024, \aap, 683, L2, \dodoi{10.1051/0004-6361/202348238}

\bibitem[{{Hu}(2021)}]{Hu2021a}
{Hu}, R. 2021, \apj, 921, 27, \dodoi{10.3847/1538-4357/ac1789}

\bibitem[{{Hu} {et~al.}(2021){Hu}, {Damiano}, {Scheucher}, {Kite}, {Seager}, \& {Rauer}}]{Hu2021}
{Hu}, R., {Damiano}, M., {Scheucher}, M., {et~al.} 2021, \apjl, 921, L8, \dodoi{10.3847/2041-8213/ac1f92}

\bibitem[{Hubeny(2017)}]{hubeny_model_2017}
Hubeny, I. 2017, Monthly Notices of the Royal Astronomical Society, 469, 841, \dodoi{10.1093/mnras/stx758}

\bibitem[{Iacono-Marziano {et~al.}(2012)Iacono-Marziano, Morizet, Le~Trong, \& Gaillard}]{iacono-marziano_new_2012}
Iacono-Marziano, G., Morizet, Y., Le~Trong, E., \& Gaillard, F. 2012, Geochimica et Cosmochimica Acta, 97, 1, \dodoi{10.1016/j.gca.2012.08.035}

\bibitem[{Izidoro {et~al.}(2021)Izidoro, Bitsch, Raymond, Johansen, Morbidelli, Lambrechts, \& Jacobson}]{izidoro_formation_2021}
Izidoro, A., Bitsch, B., Raymond, S.~N., {et~al.} 2021, Astronomy \& Astrophysics, 650, A152, \dodoi{10.1051/0004-6361/201935336}

\bibitem[{Jin \& Mordasini(2018)}]{jin_compositional_2018}
Jin, S., \& Mordasini, C. 2018, The Astrophysical Journal, 853, 163, \dodoi{10.3847/1538-4357/aa9f1e}

\bibitem[{Jin {et~al.}(2014)Jin, Mordasini, Parmentier, Van~Boekel, Henning, \& Ji}]{jin_planetary_2014}
Jin, S., Mordasini, C., Parmentier, V., {et~al.} 2014, The Astrophysical Journal, 795, 65, \dodoi{10.1088/0004-637X/795/1/65}

\bibitem[{Kite {et~al.}(2016)Kite, {Bruce Fegley Jr.}, Schaefer, \& Gaidos}]{kite_atmosphere-interior_2016}
Kite, E.~S., {Bruce Fegley Jr.}, Schaefer, L., \& Gaidos, E. 2016, The Astrophysical Journal, 828, 80, \dodoi{10.3847/0004-637X/828/2/80}

\bibitem[{Kite {et~al.}(2019)Kite, Jr, Schaefer, \& Ford}]{kite_superabundance_2019}
Kite, E.~S., Jr, B.~F., Schaefer, L., \& Ford, E.~B. 2019, The Astrophysical Journal Letters, 887, L33, \dodoi{10.3847/2041-8213/ab59d9}

\bibitem[{Kite {et~al.}(2020)Kite, Jr, Schaefer, \& Ford}]{kite_atmosphere_2020}
---. 2020, The Astrophysical Journal, 891, 111, \dodoi{10.3847/1538-4357/ab6ffb}

\bibitem[{Kite \& Schaefer(2021)}]{kite_water_2021}
Kite, E.~S., \& Schaefer, L. 2021, The Astrophysical Journal Letters, 909, L22, \dodoi{10.3847/2041-8213/abe7dc}

\bibitem[{{Lebrun} {et~al.}(2013){Lebrun}, {Massol}, {Chassefi{\`e}Re}, {Davaille}, {Marcq}, {Sarda}, {Leblanc}, \& {Brandeis}}]{Lebrun2013}
{Lebrun}, T., {Massol}, H., {Chassefi{\`e}Re}, E., {et~al.} 2013, Journal of Geophysical Research (Planets), 118, 1155, \dodoi{10.1002/jgre.20068}

\bibitem[{{Leconte} {et~al.}(2024){Leconte}, {Spiga}, {Cl{\'e}ment}, {Guerlet}, {Selsis}, {Milcareck}, {Cavali{\'e}}, {Moreno}, {Lellouch}, {Carri{\'o}n-Gonz{\'a}lez}, {Charnay}, \& {Lef{\`e}vre}}]{leconte_3d_2024}
{Leconte}, J., {Spiga}, A., {Cl{\'e}ment}, N., {et~al.} 2024, \aap, 686, A131, \dodoi{10.1051/0004-6361/202348928}

\bibitem[{{Lee} {et~al.}(2004){Lee}, {O'Neill}, {Panero}, {Shim}, {Benedetti}, \& {Jeanloz}}]{Lee2004}
{Lee}, K. K.~M., {O'Neill}, B., {Panero}, W.~R., {et~al.} 2004, Earth and Planetary Science Letters, 223, 381, \dodoi{10.1016/j.epsl.2004.04.033}

\bibitem[{Lesne {et~al.}(2015)Lesne, Scaillet, \& Pichavant}]{lesne_solubility_2015}
Lesne, P., Scaillet, B., \& Pichavant, M. 2015, Chemical Geology, 418, 104, \dodoi{10.1016/j.chemgeo.2015.03.025}

\bibitem[{{Li} {et~al.}(2015){Li}, {Gordon}, {Rothman}, {Tan}, {Hu}, {Kassi}, {Campargue}, \& {Medvedev}}]{Li2015}
{Li}, G., {Gordon}, I.~E., {Rothman}, L.~S., {et~al.} 2015, \apjs, 216, 15, \dodoi{10.1088/0067-0049/216/1/15}

\bibitem[{Libourel {et~al.}(2003)Libourel, Marty, \& Humbert}]{libourel_nitrogen_2003}
Libourel, G., Marty, B., \& Humbert, F. 2003, Geochimica et Cosmochimica Acta, 67, 4123, \dodoi{10.1016/S0016-7037(03)00259-X}

\bibitem[{Lichtenberg {et~al.}(2021)Lichtenberg, Bower, Hammond, Boukrouche, Sanan, Tsai, \& Pierrehumbert}]{lichtenberg_vertically_2021}
Lichtenberg, T., Bower, D.~J., Hammond, M., {et~al.} 2021, Journal of Geophysical Research: Planets, 126, e2020JE006711, \dodoi{10.1029/2020JE006711}

\bibitem[{Lopez \& Fortney(2013)}]{lopez_role_2013}
Lopez, E.~D., \& Fortney, J.~J. 2013, The Astrophysical Journal, 776, 2, \dodoi{10.1088/0004-637X/776/1/2}

\bibitem[{{Madhusudhan}(2018)}]{madhusudhan_review_2018}
{Madhusudhan}, N. 2018, in Handbook of Exoplanets, ed. H.~J. {Deeg} \& J.~A. {Belmonte} (Springer Cham), 104, \dodoi{10.1007/978-3-319-55333-7_104}

\bibitem[{Madhusudhan {et~al.}(2023{\natexlab{a}})Madhusudhan, Moses, Rigby, \& Barrier}]{madhusudhan_chemical_2023}
Madhusudhan, N., Moses, J.~I., Rigby, F., \& Barrier, E. 2023{\natexlab{a}}, Faraday Discussions, \dodoi{10.1039/d3fd00075c}

\bibitem[{Madhusudhan {et~al.}(2020)Madhusudhan, Nixon, Welbanks, Piette, \& Booth}]{madhusudhan_interior_2020}
Madhusudhan, N., Nixon, M.~C., Welbanks, L., Piette, A. A.~A., \& Booth, R.~A. 2020, The Astrophysical Journal Letters, 891, L7, \dodoi{10.3847/2041-8213/ab7229}

\bibitem[{Madhusudhan {et~al.}(2021)Madhusudhan, Piette, \& Constantinou}]{madhusudhan_habitability_2021}
Madhusudhan, N., Piette, A. A.~A., \& Constantinou, S. 2021, The Astrophysical Journal, 918, 1, \dodoi{10.3847/1538-4357/abfd9c}

\bibitem[{Madhusudhan {et~al.}(2023{\natexlab{b}})Madhusudhan, Sarkar, Constantinou, Holmberg, Piette, \& Moses}]{madhusudhan_carbon-bearing_2023}
Madhusudhan, N., Sarkar, S., Constantinou, S., {et~al.} 2023{\natexlab{b}}, The Astrophysical Journal Letters, 956, L13, \dodoi{10.3847/2041-8213/acf577}

\bibitem[{Mallik {et~al.}(2018)Mallik, Li, \& Wiedenbeck}]{mallik_nitrogen_2018}
Mallik, A., Li, Y., \& Wiedenbeck, M. 2018, Earth and Planetary Science Letters, 482, 556, \dodoi{10.1016/j.epsl.2017.11.045}

\bibitem[{{Matsui} \& {Abe}(1986)}]{Matsui1986}
{Matsui}, T., \& {Abe}, Y. 1986, \nat, 319, 303, \dodoi{10.1038/319303a0}

\bibitem[{Misener {et~al.}(2023)Misener, Schlichting, \& Young}]{misener_atmospheres_2023}
Misener, W., Schlichting, H.~E., \& Young, E.~D. 2023, Monthly Notices of the Royal Astronomical Society, 524, 981, \dodoi{10.1093/mnras/stad1910}

\bibitem[{Miyazaki {et~al.}(2004)Miyazaki, Hiyagon, Sugiura, Hirose, \& Takahashi}]{miyazaki_solubilities_2004}
Miyazaki, A., Hiyagon, H., Sugiura, N., Hirose, K., \& Takahashi, E. 2004, Geochimica et Cosmochimica Acta, 68, 387, \dodoi{10.1016/S0016-7037(03)00484-8}

\bibitem[{{Monteux} {et~al.}(2016){Monteux}, {Andrault}, \& {Samuel}}]{Monteux2016}
{Monteux}, J., {Andrault}, D., \& {Samuel}, H. 2016, Earth and Planetary Science Letters, 448, 140, \dodoi{10.1016/j.epsl.2016.05.010}

\bibitem[{Moore {et~al.}(1995)Moore, Vennemann, \& Carmichael}]{Moore1995}
Moore, G., Vennemann, T., \& Carmichael, I. S.~E. 1995, Geology, 23, 1099, \dodoi{10.1130/0091-7613(1995)023<1099:SOWIMT>2.3.CO;2}

\bibitem[{Ni {et~al.}(2017)Ni, Zhang, Xiong, Mao, \& Wang}]{ni_supercritical_2017}
Ni, H., Zhang, L., Xiong, X., Mao, Z., \& Wang, J. 2017, Earth-Science Reviews, 167, 62, \dodoi{10.1016/j.earscirev.2017.02.006}

\bibitem[{{Nixon} \& {Madhusudhan}(2021)}]{Nixon2021}
{Nixon}, M.~C., \& {Madhusudhan}, N. 2021, \mnras, 505, 3414, \dodoi{10.1093/mnras/stab1500}

\bibitem[{O'Neill \& Mavrogenes(2002)}]{oneill_sulfide_2002}
O'Neill, H., \& Mavrogenes, J. 2002, Journal of Petrology, 43, 1049

\bibitem[{{Orton} {et~al.}(2007){Orton}, {Gustafsson}, {Burgdorf}, \& {Meadows}}]{Orton2007}
{Orton}, G.~S., {Gustafsson}, M., {Burgdorf}, M., \& {Meadows}, V. 2007, \icarus, 189, 544, \dodoi{10.1016/j.icarus.2007.02.003}

\bibitem[{Owen \& Wu(2017)}]{owen_evaporation_2017}
Owen, J.~E., \& Wu, Y. 2017, The Astrophysical Journal, 847, 29, \dodoi{10.3847/1538-4357/aa890a}

\bibitem[{Pan {et~al.}(1991)Pan, Holloway, \& Hervig}]{pan_pressure_1991}
Pan, V., Holloway, J.~R., \& Hervig, R.~L. 1991, Geochimica et Cosmochimica Acta, 55, 1587, \dodoi{10.1016/0016-7037(91)90130-W}

\bibitem[{Papale {et~al.}(2006)Papale, Moretti, \& Barbato}]{papale_compositional_2006}
Papale, P., Moretti, R., \& Barbato, D. 2006, Chemical Geology, 229, 78, \dodoi{10.1016/j.chemgeo.2006.01.013}

\bibitem[{{Peacock} {et~al.}(2020){Peacock}, {Barman}, {Shkolnik}, {Loyd}, {Schneider}, {Pagano}, \& {Meadows}}]{Peacock2020}
{Peacock}, S., {Barman}, T., {Shkolnik}, E.~L., {et~al.} 2020, \apj, 895, 5, \dodoi{10.3847/1538-4357/ab893a}

\bibitem[{Piette \& Madhusudhan(2020)}]{piette_temperature_2020}
Piette, A. A.~A., \& Madhusudhan, N. 2020, The Astrophysical Journal, 904, 154, \dodoi{10.3847/1538-4357/abbfb1}

\bibitem[{Richard {et~al.}(2012)Richard, Gordon, Rothman, Abel, Frommhold, Gustafsson, Hartmann, Hermans, Lafferty, Orton, Smith, \& Tran}]{Richard2012}
Richard, C., Gordon, I., Rothman, L., {et~al.} 2012, \jqsrt, 113, 1276 , \dodoi{https://doi.org/10.1016/j.jqsrt.2011.11.004}

\bibitem[{Rigby \& Madhusudhan(2024)}]{rigby_ocean_2024}
Rigby, F.~E., \& Madhusudhan, N. 2024, Monthly Notices of the Royal Astronomical Society, 529, 409, \dodoi{10.1093/mnras/stae413}

\bibitem[{Rogers {et~al.}(2021)Rogers, Gupta, Owen, \& Schlichting}]{rogers_photoevaporation_2021}
Rogers, J.~G., Gupta, A., Owen, J.~E., \& Schlichting, H.~E. 2021, Monthly Notices of the Royal Astronomical Society, 508, 5886, \dodoi{10.1093/mnras/stab2897}

\bibitem[{{Rogers} {et~al.}(2011){Rogers}, {Bodenheimer}, {Lissauer}, \& {Seager}}]{Rogers2011}
{Rogers}, L.~A., {Bodenheimer}, P., {Lissauer}, J.~J., \& {Seager}, S. 2011, \apj, 738, 59, \dodoi{10.1088/0004-637X/738/1/59}

\bibitem[{Roskosz {et~al.}(2013)Roskosz, Bouhifd, Jephcoat, Marty, \& Mysen}]{roskosz_nitrogen_2013}
Roskosz, M., Bouhifd, M., Jephcoat, A., Marty, B., \& Mysen, B. 2013, Geochimica et Cosmochimica Acta, 121, 15, \dodoi{10.1016/j.gca.2013.07.007}

\bibitem[{Rothman {et~al.}(2010)Rothman, Gordon, Barber, Dothe, Gamache, Goldman, Perevalov, Tashkun, \& Tennyson}]{Rothman2010}
Rothman, L., Gordon, I., Barber, R., {et~al.} 2010, \jqsrt, 111, 2139 , \dodoi{https://doi.org/10.1016/j.jqsrt.2010.05.001}

\bibitem[{Schaefer \& Fegley(2017)}]{Schaefer2017}
Schaefer, L., \& Fegley, B. 2017, The Astrophysical Journal, 843, 120, \dodoi{10.3847/1538-4357/aa784f}

\bibitem[{Schaefer {et~al.}(2016)Schaefer, Wordsworth, Berta-Thompson, \& Sasselov}]{Schaefer2016}
Schaefer, L., Wordsworth, R.~D., Berta-Thompson, Z., \& Sasselov, D. 2016, The Astrophysical Journal, 829, 63, \dodoi{10.3847/0004-637X/829/2/63}

\bibitem[{Schlichting \& Young(2022)}]{schlichting_chemical_2022}
Schlichting, H.~E., \& Young, E.~D. 2022, The Planetary Science Journal, 3, 127, \dodoi{10.3847/PSJ/ac68e6}

\bibitem[{{Seager} {et~al.}(2007){Seager}, {Kuchner}, {Hier-Majumder}, \& {Militzer}}]{Seager2007}
{Seager}, S., {Kuchner}, M., {Hier-Majumder}, C.~A., \& {Militzer}, B. 2007, \apj, 669, 1279, \dodoi{10.1086/521346}

\bibitem[{{Shorttle} {et~al.}(2024){Shorttle}, {Jordan}, {Nicholls}, {Lichtenberg}, \& {Bower}}]{shorttle_distinguishing_2024}
{Shorttle}, O., {Jordan}, S., {Nicholls}, H., {Lichtenberg}, T., \& {Bower}, D.~J. 2024, \apjl, 962, L8, \dodoi{10.3847/2041-8213/ad206e}

\bibitem[{Silver {et~al.}(1990)Silver, Ihinger, \& Stolper}]{silver_influence_1990}
Silver, L.~A., Ihinger, P.~D., \& Stolper, E. 1990, Contributions to Mineralogy and Petrology, 104, 142, \dodoi{10.1007/BF00306439}

\bibitem[{Sossi {et~al.}(2023)Sossi, Tollan, Badro, \& Bower}]{sossi_solubility_2023}
Sossi, P.~A., Tollan, P. M.~E., Badro, J., \& Bower, D.~J. 2023, Earth and Planetary Science Letters, 601, 117894, \dodoi{10.1016/j.epsl.2022.117894}

\bibitem[{{Stock} {et~al.}(2022){Stock}, {Kitzmann}, \& {Patzer}}]{Stock2022}
{Stock}, J.~W., {Kitzmann}, D., \& {Patzer}, A. B.~C. 2022, \mnras, 517, 4070, \dodoi{10.1093/mnras/stac2623}

\bibitem[{{Stock} {et~al.}(2018){Stock}, {Kitzmann}, {Patzer}, \& {Sedlmayr}}]{Stock2018}
{Stock}, J.~W., {Kitzmann}, D., {Patzer}, A. B.~C., \& {Sedlmayr}, E. 2018, \mnras, 479, 865, \dodoi{10.1093/mnras/sty1531}

\bibitem[{{Stolper}(1982)}]{Stolper1982}
{Stolper}, E. 1982, \gca, 46, 2609, \dodoi{10.1016/0016-7037(82)90381-7}

\bibitem[{Suer {et~al.}(2023)Suer, Jackson, Grewal, Dalou, \& Lichtenberg}]{suer_distribution_2023}
Suer, T.-A., Jackson, C., Grewal, D.~S., Dalou, C., \& Lichtenberg, T. 2023, Frontiers in Earth Science, 11, 1159412, \dodoi{10.3389/feart.2023.1159412}

\bibitem[{{Tashkun} {et~al.}(2015){Tashkun}, {Perevalov}, {Gamache}, \& {Lamouroux}}]{Tashkun2015}
{Tashkun}, S.~A., {Perevalov}, V.~I., {Gamache}, R.~R., \& {Lamouroux}, J. 2015, \jqsrt, 152, 45, \dodoi{10.1016/j.jqsrt.2014.10.017}

\bibitem[{{Thomas} \& {Asimow}(2013)}]{Thomas2013}
{Thomas}, C.~W., \& {Asimow}, P.~D. 2013, Journal of Geophysical Research (Solid Earth), 118, 5738, \dodoi{10.1002/jgrb.50374}

\bibitem[{{Tian} \& {Heng}(2024)}]{tian_atmospheric_2024}
{Tian}, M., \& {Heng}, K. 2024, \apj, 963, 157, \dodoi{10.3847/1538-4357/ad217c}

\bibitem[{{Tsai} {et~al.}(2021){Tsai}, {Innes}, {Lichtenberg}, {Taylor}, {Malik}, {Chubb}, \& {Pierrehumbert}}]{Tsai2021}
{Tsai}, S.-M., {Innes}, H., {Lichtenberg}, T., {et~al.} 2021, \apjl, 922, L27, \dodoi{10.3847/2041-8213/ac399a}

\bibitem[{{Underwood} {et~al.}(2016){Underwood}, {Tennyson}, {Yurchenko}, {Huang}, {Schwenke}, {Lee}, {Clausen}, \& {Fateev}}]{Underwood2016}
{Underwood}, D.~S., {Tennyson}, J., {Yurchenko}, S.~N., {et~al.} 2016, \mnras, 459, 3890, \dodoi{10.1093/mnras/stw849}

\bibitem[{{Valencia} {et~al.}(2013){Valencia}, {Guillot}, {Parmentier}, \& {Freedman}}]{Valencia2013}
{Valencia}, D., {Guillot}, T., {Parmentier}, V., \& {Freedman}, R.~S. 2013, \apj, 775, 10, \dodoi{10.1088/0004-637X/775/1/10}

\bibitem[{Venturini {et~al.}(2020)Venturini, Guilera, Haldemann, Ronco, \& Mordasini}]{venturini_nature_2020}
Venturini, J., Guilera, O.~M., Haldemann, J., Ronco, M.~P., \& Mordasini, C. 2020, Astronomy \& Astrophysics, 643, L1, \dodoi{10.1051/0004-6361/202039141}

\bibitem[{{Venturini} {et~al.}(2024){Venturini}, {Ronco}, {Guilera}, {Haldemann}, {Mordasini}, \& {Miller Bertolami}}]{venturini_fading_2024}
{Venturini}, J., {Ronco}, M.~P., {Guilera}, O.~M., {et~al.} 2024, \aap, 686, L9, \dodoi{10.1051/0004-6361/202349088}

\bibitem[{{Wogan} {et~al.}(2024){Wogan}, {Batalha}, {Zahnle}, {Krissansen-Totton}, {Tsai}, \& {Hu}}]{wogan_jwst_2024}
{Wogan}, N.~F., {Batalha}, N.~E., {Zahnle}, K.~J., {et~al.} 2024, \apjl, 963, L7, \dodoi{10.3847/2041-8213/ad2616}

\bibitem[{{Woitke} {et~al.}(2018){Woitke}, {Helling}, {Hunter}, {Millard}, {Turner}, {Worters}, {Blecic}, \& {Stock}}]{Woitke2018}
{Woitke}, P., {Helling}, C., {Hunter}, G.~H., {et~al.} 2018, \aap, 614, A1, \dodoi{10.1051/0004-6361/201732193}

\bibitem[{Woodland {et~al.}(2019)Woodland, Girnis, Bulatov, Brey, \& Höfer}]{woodland_experimental_2019}
Woodland, A.~B., Girnis, A.~V., Bulatov, V.~K., Brey, G.~P., \& Höfer, H.~E. 2019, Chemical Geology, 505, 12, \dodoi{10.1016/j.chemgeo.2018.12.008}

\bibitem[{{Wordsworth}(2016)}]{Wordsworth2016}
{Wordsworth}, R.~D. 2016, Earth and Planetary Science Letters, 447, 103, \dodoi{10.1016/j.epsl.2016.04.002}

\bibitem[{Yoshioka {et~al.}(2019)Yoshioka, Nakashima, Nakamura, Shcheka, \& Keppler}]{yoshioka_carbon_2019}
Yoshioka, T., Nakashima, D., Nakamura, T., Shcheka, S., \& Keppler, H. 2019, Geochimica et Cosmochimica Acta, 259, 129, \dodoi{10.1016/j.gca.2019.06.007}

\bibitem[{{Yu} {et~al.}(2021){Yu}, {Moses}, {Fortney}, \& {Zhang}}]{Yu2021}
{Yu}, X., {Moses}, J.~I., {Fortney}, J.~J., \& {Zhang}, X. 2021, \apj, 914, 38, \dodoi{10.3847/1538-4357/abfdc7}

\bibitem[{{Yurchenko} {et~al.}(2011){Yurchenko}, {Barber}, \& {Tennyson}}]{Yurchenko2011}
{Yurchenko}, S.~N., {Barber}, R.~J., \& {Tennyson}, J. 2011, \mnras, 413, 1828, \dodoi{10.1111/j.1365-2966.2011.18261.x}

\bibitem[{Yurchenko \& Tennyson(2014)}]{Yurchenko2014}
Yurchenko, S.~N., \& Tennyson, J. 2014, \mnras, 440, 1649, \dodoi{10.1093/mnras/stu326}

\bibitem[{Zajacz {et~al.}(2013)Zajacz, Candela, Piccoli, Sanchez-Valle, \& Wälle}]{zajacz_solubility_2013}
Zajacz, Z., Candela, P.~A., Piccoli, P.~M., Sanchez-Valle, C., \& Wälle, M. 2013, Geochimica et Cosmochimica Acta, 112, 288, \dodoi{10.1016/j.gca.2013.02.026}

\bibitem[{Zeng {et~al.}(2019)Zeng, Jacobsen, Sasselov, Petaev, Vanderburg, Lopez-Morales, Perez-Mercader, Mattsson, Li, Heising, Bonomo, Damasso, Berger, Cao, Levi, \& Wordsworth}]{zeng_growth_2019}
Zeng, L., Jacobsen, S.~B., Sasselov, D.~D., {et~al.} 2019, Proceedings of the National Academy of Science, 116, 9723, \dodoi{10.1073/pnas.1812905116}

\end{thebibliography}
\bibliographystyle{aasjournal}
\FloatBarrier

\appendix

\section{Availability of Solubility Laws} \label{app:solubility}
We discuss here the availability of silicate melt solubility laws for the volatile species of interest, at the chemical and physical conditions relevant for magma oceans on temperate sub-Neptunes in the gas dwarf scenario. These findings motivate our choices for the solubility laws adopted in this work. We compile a bibliography of the solubility laws consulted for the preparation of this work in Table~\ref{tab:solub_list}, and show a selection of them in Figure~\ref{fig:solubilities}. For most composition-dependent laws, we adopt a basalt composition for the melt -- specifically, when a law explicitly depends on melt composition parameters, we set these to the values corresponding to the Mt Etna basalt from \citet{iacono-marziano_new_2012}. This choice is due to the wide availability of solubility laws for basaltic melt and because of the association of basalt with peridotite, which we assume to be the mantle composition. 

\subsection{Nitrogen Species}
The solubility of \ch{N2} has been explored for a wide range of parameters (e.g., \citealp{libourel_nitrogen_2003, miyazaki_solubilities_2004, roskosz_nitrogen_2013,  mallik_nitrogen_2018,  boulliung_oxygen_2020, bernadou_nitrogen_2021, gao_experimental_2022}), at pressures up to 14.8 GPa and temperatures up to 2800 K \citep{roskosz_nitrogen_2013}. By compiling the available data at $P \leq 8.2$ GPa and adding their own measurements, \citet{dasgupta_fate_2022} proposed the solubility law which we use in our calculations.
This law, however, does not appear to extrapolate well at higher pressures and moderately reduced conditions ($f_{\ch{O2}} \sim \mathrm{IW} - 2$). As warned by \citet{dasgupta_fate_2022}, experimental data indicate the solubility seems to reach a plateau, while the law predicts solubility to monotonically increase with pressure. A direct comparison with \citet{roskosz_nitrogen_2013}'s $10$ GPa and $14.8$ GPa data points reveals indeed a true solubility $\sim 1$ order of magnitude lower than predicted using \citet{dasgupta_fate_2022}'s law for pure nitrogen vapor. At the extremely reduced conditions explored here, the plateau effect is expected to already be significant at lower pressures \citep{dasgupta_fate_2022}.

It is also noteworthy that \citet{gao_experimental_2022} - whose data was included in the \citet{dasgupta_fate_2022} dataset - find some indication of a decrease in the \textit{physical} solubility of \ch{N2} already at $P=8$ GPa. We note that physical solubility is expected to be the dominant solubility mechanism at the oxidized conditions explored by \citealt{gao_experimental_2022}, as opposed to the chemical solubility relevant at reduced conditions \citep{libourel_nitrogen_2003}.
Nevertheless, as the relevant quantity is not the total pressure, but rather the nitrogen partial pressure, we believe that the \citet{dasgupta_fate_2022} law can still be a reasonably good approximation even at the reducing, high-pressure conditions that apply at the magma-atmosphere interface, given that the expected \ch{N2} mixing ratio in the atmosphere is $\lesssim 10^{-4}$ in the present models.

The lack of data or simulations for the solubility of \ch{NH3} in silicate melt leads us to neglect it, with the caveat that this will lead to our calculations setting only an upper limit on the abundance of \ch{N}-bearing species in the atmosphere.
\subsection{Carbon Species}
Of the three prominent carbon species (\ch{CO2}, \ch{CO}, \ch{CH4}), \ch{CO2} is by far the one for which the most complete experimental data on solubility in magma is available (e.g., \citealp{pan_pressure_1991,dixon_experimental_1995, papale_compositional_2006, iacono-marziano_new_2012}). 
Considering this wide dataset, for the case of $T = 2273$ K and a bulk silicate Earth (BSE) melt composition, \citet{suer_distribution_2023} find that the solubility of \ch{CO2} is well-approximated by Henry's Law, which they fit to the data. \citet{suer_distribution_2023}'s law is in excellent agreement with high pressure ($P \geq 8$ GPa) molecular dynamics simulations by \citet{guillot_carbon_2011} for the $T = 2273$ K and rhyolite case, and in good agreement with the corresponding mid-ocean ridge basalt (MORB) case. Interestingly, the agreement is slightly worse with the kimberlite melt case, where instead the melt is closest to Suer's BSE composition. 

At lower pressures, the agreement with the simulations is worse, but still within a factor of order unity. In any case, the agreement between \citet{suer_distribution_2023}'s law and \citet{guillot_carbon_2011}'s simulations is always satisfactory, which also highlights the weak dependence of the solubility of \ch{CO2} on the melt composition, particularly at $P \leq 8$ GPa \citep{guillot_carbon_2011}. 
Despite the fact that the \citet{suer_distribution_2023} law is intended for lower temperatures than those relevant in this study, due to the lack of more appropriate alternatives we adopt it in our calculations.\\

For \ch{CO}, there is a lack of solubility data at high pressure and temperature. Solubility laws are provided by, e.g., \citet{armstrong_speciation_2015} and \citet{yoshioka_carbon_2019} (for both MORB and rhyolite melts), both of whom carried out experiments at $P \sim $ 1 GPa and $T \sim 1500 ^\circ$C. 
The lack of data may be explained by the fact that exploring the solubility of \ch{CO} at high pressures is especially complicated, because the $2\ch{CO} = \ch{C} + \ch{CO2}$ reaction gets skewed to the right as pressure grows, making an initially pure \ch{CO} vapour spontaneously become mostly \ch{CO2} at $P \gtrsim 1$ GPa \citep{yoshioka_carbon_2019}. 

An alternative prescription, used by \citet{schlichting_chemical_2022} as informed by \citet{hirschmann_constraints_2016}, is to instead set the solubility of \ch{CO} to be one third of that of \ch{CO2}. This method, taking \citet{suer_distribution_2023}'s BSE law for the \ch{CO2} solubility, yields a \ch{CO} solubility significantly higher than any of the other laws mentioned so far. This might be due to \citet{suer_distribution_2023}'s law being tested for different temperatures and melt compositions. This, however, seems unlikely: on the one hand, the solubility of \ch{CO} is only weakly dependent on temperature \citep{yoshioka_carbon_2019}; on the other hand, \citet{yoshioka_carbon_2019}'s two laws for MORB and rhyolite yield results less than an order of magnitude apart. This indicates a comparatively weak dependence of solubility on melt composition, while \citet{schlichting_chemical_2022}'s prescription applied to \citet{suer_distribution_2023}'s law results in a solubility between 1 and 2 orders of magnitude higher, depending on the pressure.
Ultimately, we use the \citet{yoshioka_carbon_2019} MORB law in our calculations, due to it being more recent and calibrated at higher pressures than \citet{armstrong_speciation_2015}.\\

The data on \ch{CH4} seems to be even sparser: the only solubility law we encountered in the literature -- which we use here -- is the one in \citet{ardia_solubility_2013}, resulting from experiments at $0.7 \leq P \leq 3$ GPa and $1400 \leq T \leq 1450 ^\circ$C, and the Henry's Law fit to their data by \citet{lichtenberg_vertically_2021} - the law by \citet{ardia_solubility_2013}, indeed, follows Henry's law for total pressures $P \lesssim 10^4$ bar (at $T=3000$ K, regardless of the \ch{CH4} partial pressure).\\

\subsection{Other Volatiles}
The other major volatiles of note are expected to be sulfur species, water, and \ch{H2}, as well as \ch{He}, which, however, as a noble gas, has no impact on the atmospheric chemistry.

Chemical equilibrium calculations indicate that, at conditions relevant at the interface, sulfur will be mostly in \ch{H2S}, with little \ch{S2}. 
For \ch{S2} we use the law by \citet{gaillard_redox_2022}.
It should be noted that this law is calibrated only with data collected at atmospheric pressure, relatively low temperature ($T \leq 1673$ K), and not very reducing conditions ($\Delta \mathrm{IW} \geq -1$). As such, its extrapolation to the extreme conditions explored in this paper should be considered only as a zeroth-order estimate of the true solubility. No significant high-pressure/high-temperature data to compare with \citet{gaillard_redox_2022}'s predictions were found either, the \citet{woodland_experimental_2019} high-pressure data being for a \textit{carbonate}-silicate melt.

For \ch{H2S}, we found two laws in the literature, by \citet{clemente_solubility_2004} -- for rhyolite -- and by \citet{lesne_solubility_2015} -- for basaltic melts. The former is calibrated for $1073 \leq T \leq 1273$ K and $P = 2 \times 10^3$ bar, while the latter for $1323 \leq T \leq 1473$ K and $250 \leq P \leq  2 \times 10^3$ bar. The two laws differ significantly in their temperature dependence: \citet{clemente_solubility_2004} find that the solubility of \ch{H2S} moderately increases with increasing temperature, while the law by \citet{lesne_solubility_2015} indicates an extremely strong and negative temperature dependence. Furthermore, \citet{lesne_solubility_2015} include a dependence on the mole fraction of \ch{FeO} in the magma, while the law by \citet{clemente_solubility_2004} only depends on thermodynamic parameters. 
However, when extrapolated to high temperature ($T \sim 3000$ K) and pressure ($P \sim 10^5$ bar), both laws predict negligibly small solubility for \ch{H2S} at the expected mixing ratios (shown in Figure~\ref{fig:melt}). Hence, we do not expect the results of our investigation to be noticeably impacted by the choice of one law over the other. In this investigation, we chose to use the law by \citet{clemente_solubility_2004}.

For \ch{H2}, the law most used in the literature we reviewed is by \citet{hirschmann_solubility_2012}, who carry out experiments at $0.7 \leq P \leq 3$ GPa and $1400 \leq T \leq 1500 ^\circ$C, and give two expressions, for basaltic and andesitic melt. Their law for basaltic melt is in excellent agreement with that given by \citet{suer_distribution_2023} for BSE melt up to $P \sim 1$ GPa, and so is, to a slightly lesser extent, their andesitic melt law. At higher pressures, however, they diverge, with \citet{suer_distribution_2023} predicting Henrian behaviour to arbitrary pressure, while \citet{hirschmann_solubility_2012}'s laws predict a decline in solubility as pressure increases, consistently with their experimental results. 

As the laws given in \citet{hirschmann_solubility_2012} have a robust high-pressure experimental background, we use those, and here, specifically, the basaltic melt case.\\ 

For \ch{H2O}, there is a great deal of experimental data on the solubility in silicate melts \citep[e.g.,][]{Stolper1982,silver_influence_1990, Moore1995, dixon_experimental_1995, papale_compositional_2006, iacono-marziano_new_2012, sossi_solubility_2023}, a complete review of which is beyond the scope of this work. We focus here on two solubility laws: \citet{sossi_solubility_2023}, the most recent law available, and \citet{iacono-marziano_new_2012}, which is the one we choose to implement in our study. 
\citet{sossi_solubility_2023} provide two slightly different estimates depending on the value of the molar absorption coefficient $\epsilon_{3550}$, each depending linearly on the square roots of the atmospheric fugacities of both water and molecular hydrogen. These are the result of experiments carried out at very low pressure ($P = 1$ atm) and high temperature ($T=2173$ K). 

Higher-pressure experiments are carried out in \citet{iacono-marziano_new_2012}, who also propose a solubility law, calibrated upon a vast but low-temperature experimental database ($10^2 \leq P \leq 10^4$ bar, $1100 \leq T \leq 1400 ^\circ$C), which is in rough agreement with that of \citet{sossi_solubility_2023} for an \ch{H2}-rich envelope. 

The fact that \citet{sossi_solubility_2023}'s law depends on a linear combination of the square roots of the fugacities of both \ch{H2} and \ch{H2O}, however, risks breaking element conservation for oxygen: indeed, it would predict some dissolved \ch{O} in the magma even if no \ch{O} is present -- in any species -- in the initial atmospheric composition. This effect is expected to be particularly relevant at the very reduced conditions explored here, where the abundance of \ch{O} is expected to be low. 

We thus consider extrapolating the law of \citet{iacono-marziano_new_2012} to higher temperatures to be a more accurate prescription than extrapolating that by \citet{sossi_solubility_2023} to high pressures, and hence do so here, assuming an Etna basalt composition for the melt. This choice is also consistent with that in \citet{gaillard_redox_2022}.  
\subsection{Summary}
\begin{figure}
    \centering
    \includegraphics[width=0.9\linewidth]{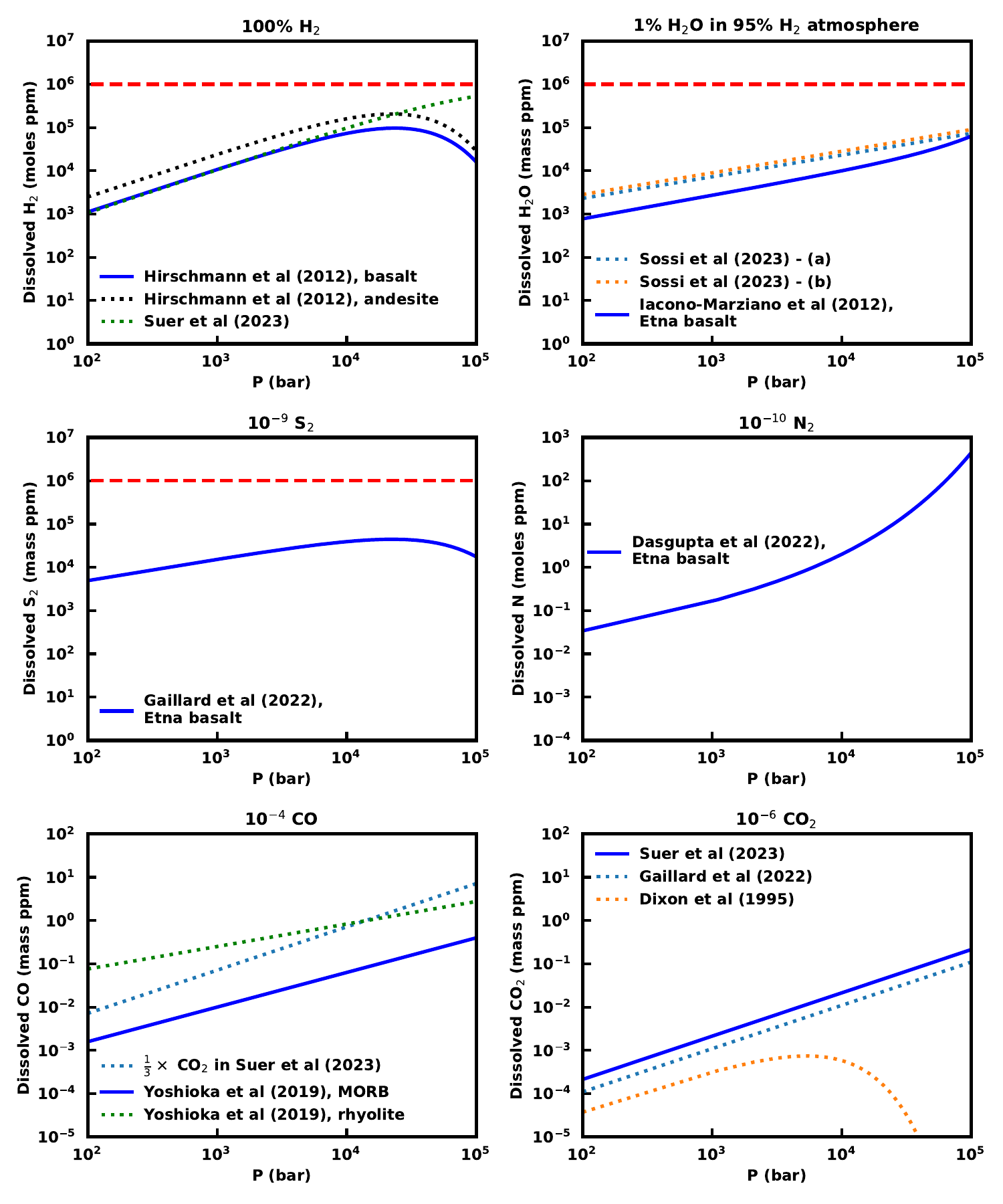}
    \caption{Behaviour of selected solubility laws in silicate melt for some prominent molecules. Solid blue lines indicate the laws used in this study. All temperature-dependent laws are shown here for $T = 3000$ K; the $\Delta$IW-dependent \ch{N2} and \ch{S2} laws are shown for $\Delta \mathrm{IW} = -6.4$; all laws with free composition-dependent parameters are shown for the  Mt. Etna basalt composition \citep{iacono-marziano_new_2012}. For \ch{H2O}, we label as (a) the law given in \citet{sossi_solubility_2023} for $\epsilon_{3550} = 6.3$ m$^2$/mol, and as (b) that for $\epsilon_{3550} = 5.1$ m$^2$/mol. The mixing ratio of each species, informed by Section~\ref{sec:results/interface}, is indicated in the title of the respective subplot; the x-axis indicates total pressure. The gases are treated ideally, i.e. we take the fugacity of each species' to be equal to its partial pressure. We do not plot the \ch{CH4} solubility, for which \citet{ardia_solubility_2013} -- whose law we use -- predicts a much lower value than the other carbon-bearing species, nor the solubility of \ch{H2S}, for which both \citet{clemente_solubility_2004} and \citet{lesne_solubility_2015} predict much lower solubility than for \ch{S2}}
    \label{fig:solubilities}
\end{figure}

Data on solubility in silicate melt are available, at some conditions, for several species of interest, with the one exception being \ch{NH3}, for which we were unable to find any solubility laws. We list the bibliography on solubility laws and/or data points we have explored for this study in Table~\ref{tab:solub_list}, and we show a selection of them in Figure~\ref{fig:solubilities}. In general, the scenario explored in this study, relevant for magma oceans on temperate gas dwarfs, is extreme in a threefold way: it leads to high temperatures ($T \gtrsim 2500$ K), high pressures ($P \gtrsim 10^5$ bar), and very reduced melts compared to Earth ($\Delta \mathrm{IW} \lesssim - 5$). There is no data, for any species, at such conditions in all three ways. Only for \ch{N2} data at both very high temperature and pressure exists, but that is for relatively oxidised conditions \citep{roskosz_nitrogen_2013}. High-pressure ($P \geq 10^5$ bar) simulations exist for \ch{CO2}, but only at $T \leq 2273$. For \ch{S2}, for a co-existing fluid phase, high-pressure data only exists at low temperature, and only for carbonate-silicate melt \citep{woodland_experimental_2019}. All other species seem to lack high-pressure data.\\ 
Exploring this region of the parameter space, either experimentally or through simulations, will be crucial for improving our understanding of potential magma oceans in sub-Neptunes, and our ability to lift observational degeneracies with other possible internal structures.
\begin{table}[]
\begin{tabular}{|l|l|l|l|}
\hline
Species             & Reference       & Temperature (K)         & Pressure (bar)                                 \\ \hline
\ch{N2}             & \citet{dasgupta_fate_2022}        & $1323 \leq T \leq 2700$ & $1 \leq P \leq 8.2 \times 10^4$                \\
\ch{N2}             & \citet{gao_experimental_2022}             & $1473 \leq T \leq 1873$ & $3 \times 10^3 \leq P \leq 8 \times 10^4$      \\

\ch{N2}             & \citet{bernadou_nitrogen_2021}      & $1473 \leq T \leq 1573$ & $8 \times 10^2 \leq P \leq 10^4$               \\
\ch{N2}             & \citet{boulliung_oxygen_2020}     & $T = 1698$              & $P=1$                                          \\
\ch{N2}             & \citet{mallik_nitrogen_2018}          & $1323 \leq T \leq 1573$ & $2 \times 10^4 \leq P  \leq 4 \times 10^4$     \\
\ch{N2}             & \citet{roskosz_nitrogen_2013}        & $2500 \leq T \leq 2800$ & $1.8 \times 10^4 \leq T \leq 1.48 \times 10^5$ \\
\ch{N2}             & \citet{miyazaki_solubilities_2004}        & $1573 \leq T \leq 1823$ & $1 \leq T \leq 2 \times 10^3$                  \\
\ch{N2}             & \citet{libourel_nitrogen_2003}        & $1673 \leq T \leq 1698$ & $P=1$                                          \\ \hline
\ch{CO2}            & \citet{suer_distribution_2023}         & $T = 2273$ & $1 \leq P \leq 10^2$             \\
\ch{CO2}            & \citet{guillot_carbon_2011}$^\mathrm{a}$         & $1473 \leq T \leq 2273$ & $10^3 \leq P \leq 1.5 \times 10^5$             \\

\ch{CO2}            & \citet{pan_pressure_1991}             & $1443 \leq T \leq 1873$ & $10^3 \leq P \leq 1.5 \times 10^4$             \\ \hline
\ch{CO2}, \ch{ H2O} & \citet{iacono-marziano_new_2012} & $1373 \leq T \leq 1673$ & $10^2 \leq P \leq 10^4$                        \\
\ch{CO2}, \ch{H2O}  & \citet{papale_compositional_2006}          & $1073 \leq T \leq 1973$ & $191 \leq P \leq 3.5 \times 10^4$              \\
\ch{CO2}, \ch{H2O}  & \citet{dixon_experimental_1995}           & $T=1473$                & $2.01 \times 10^2 \leq P \leq 9.8 \times 10^2$ \\\hline
\ch{H2O}            & \citet{sossi_solubility_2023}           & $T=2173$                & $P=1$                                          \\
\ch{H2O}            & \citet{Moore1995}          & $973 \leq T \leq 1473$ & $1 \leq P \leq 2 \times 10^3$                 \\ 
\ch{H2O}            & \citet{silver_influence_1990}          & $1123 \leq T \leq 1723$ & $49 \leq P \leq 2 \times 10^4$                 \\ 
\hline
\ch{CO}             & \citet{yoshioka_carbon_2019}        & $1473 \leq T \leq 1873$ & $2.08 \times 10^3 \leq P \leq 3 \times 10^4$   \\ 
\ch{CO}             & \citet{armstrong_speciation_2015}       & $T = 1673$              & $P = 1.2 \times 10^4$                          \\ \hline
\ch{CH4}            & \citet{ardia_solubility_2013}           & $1673 \leq T \leq 1723$ & $7 \times 10^3 \leq P \leq 3 \times 10^4$      \\ \hline

\ch{S2}             & \citet{boulliung_sulfur_2023}       & $1473 \leq T \leq 1773$ & $P=1$                                          \\
\ch{S2}             & \citet{gaillard_redox_2022}$^\mathrm{b}$        & $1073 \leq T \leq 1673$ & $P=1$                           \\ 
\ch{S2}             & \citet{woodland_experimental_2019}$^\mathrm{c}$        & $1673 \leq T \leq 1873$ & $5 \times 10^4 \leq P \leq 1.05 \times 10^5$        \\
\ch{S2}             & \citet{oneill_sulfide_2002}         & $T = 1673$              & $P = 1$                                        \\
\hline
\ch{H2S}  & \citet{lesne_solubility_2015}        & $1323 \leq T \leq 1473$ & $ 250 \leq P \leq 2 \times 10^3$                            \\ \hline
\ch{H2S}, \ch{SO2}  & \citet{clemente_solubility_2004}        & $1073 \leq T \leq 1273$ & $P = 2 \times 10^3$             \\ \hline
\ch{H2}             & \citet{hirschmann_solubility_2012}      & $1673 \leq T \leq 1773$ & $7 \times 10^3 \leq P \leq 3 \times 10^4$      \\ \hline
\end{tabular}
\caption{Sources of solubility data and laws considered in this study. The (total) pressure and temperature ranges indicated are those corresponding to the experiments carried out in the respective studies, or, if the studies calibrate a solubility law based on data from previous works, the range spanned by those.\\
$^\mathrm{a}$: Molecular dynamics simulation.\\
$^\mathrm{b}$: \citet{gaillard_redox_2022} state that their law is calibrated against data obtained with gas at a pressure of one atmosphere. However, they refer to \citet{zajacz_solubility_2013}, whose experiments were carried out at 200 MPa.\\
$^\mathrm{c}$: For carbonate-silicate melt.
}
\label{tab:solub_list}
\end{table}

\section{Sensitivity to Atmospheric Parameters} \label{app:atmosphericchem}

As described in Section~\ref{sec:results/5}, we explore a range of values for three key atmospheric parameters that could influence the observable composition: the metallicity, the eddy diffusion coefficient $K_\mathrm{zz}$, and the internal temperature $T_\mathrm{int}$. Our canonical cases, shown in Figure~\ref{fig:Xprofiles}, correspond to $P$-$T$ profiles C2 and C3 with median metallicity of $50\times$solar, with and without elemental depletion respectively, and $K_\mathrm{zz}$ of 10$^6$ cm$^2$s$^{-1}$ in the deep atmosphere. We investigate if a higher metallicity, a broader range of $K_\mathrm{zz}$ values and/or a higher $T_\mathrm{int}$ could better match the observed abundances than our canonical cases, for example with higher CO$_2$ abundance. We therefore consider models with higher metallicities of 100$\times$solar and 300$\times$solar, and two end-member scenarios of 10$^4$ cm$^2$s$^{-1}$ and 10$^8$ cm$^2$s$^{-1}$ for $K_\mathrm{zz}$ in the deep, convective region. We also consider the effect of using a higher value of $T_\mathrm{int}$ of $60$~K, as previously considered by \citet{Hu2021a}. Disequilibrium effects due to photochemistry and vertical mixing are included in all cases discussed here.

\begin{figure*}
    \centering
    \includegraphics[width=1.0\textwidth]{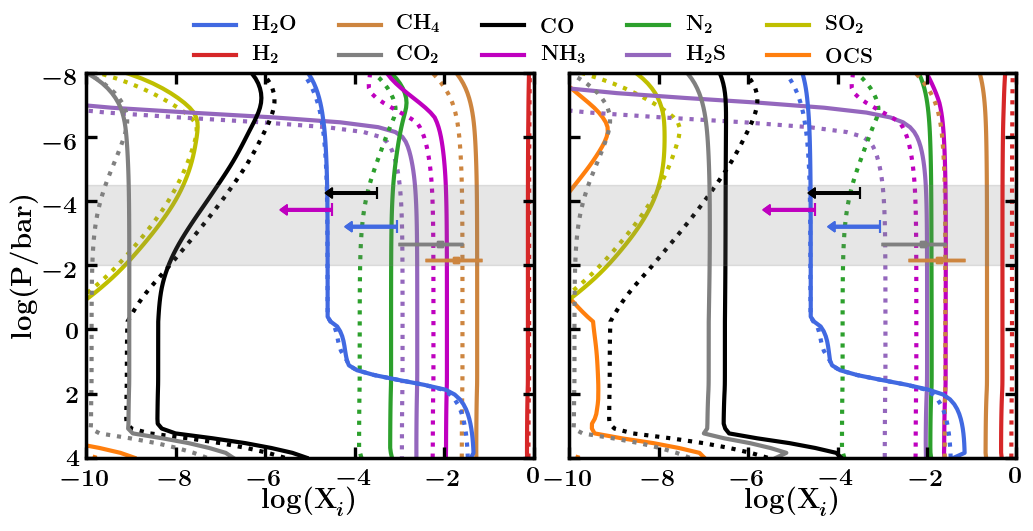}
    \caption{Effect of enhanced metallicity on the vertical mixing ratio profiles for several H-C-O-N-S molecular species with the C2 $P$-$T$ profile. Dotted lines show the profile from the left-hand side of Figure \ref{fig:Xprofiles}, for C2 with $30\%$ silicates, equivalent to $50\times$ solar elemental abundances. Left: solid lines indicate the corresponding profiles for $100\times$ solar metallicity. Right: solid lines indicate the corresponding profiles for $300\times$ solar metallicity.}
    \label{fig:Xprofiles_metallicity}
\end{figure*}

We start with investigating departures from the canonical C2 case, as shown in Figure~\ref{fig:Xprofiles}. We first fix the $K_\mathrm{zz}$ profile to that used in the canonical case and vary the metallicity as described above. The resulting vertical mixing ratio profiles are shown in Figure~\ref{fig:Xprofiles_metallicity}, along with those for $50\times$ metallicity from Figure~\ref{fig:Xprofiles} for comparison. For both the $100\times$ and $300\times$ solar cases, the abundance of CO$_2$ remains lower than that of CO throughout the atmosphere, as for the $50\times$ solar case. Similarly, the CO$_2$ and NH$_3$ abundances are inconsistent with the retrieved values in the photosphere, between $\sim$0.01-10 mbar, in all cases. Additionally, the CH$_4$ abundance for $300\times$ solar metallicity is higher than the retrieved abundance. 

\begin{figure*}
    \centering
    \includegraphics[width=0.9\textwidth]{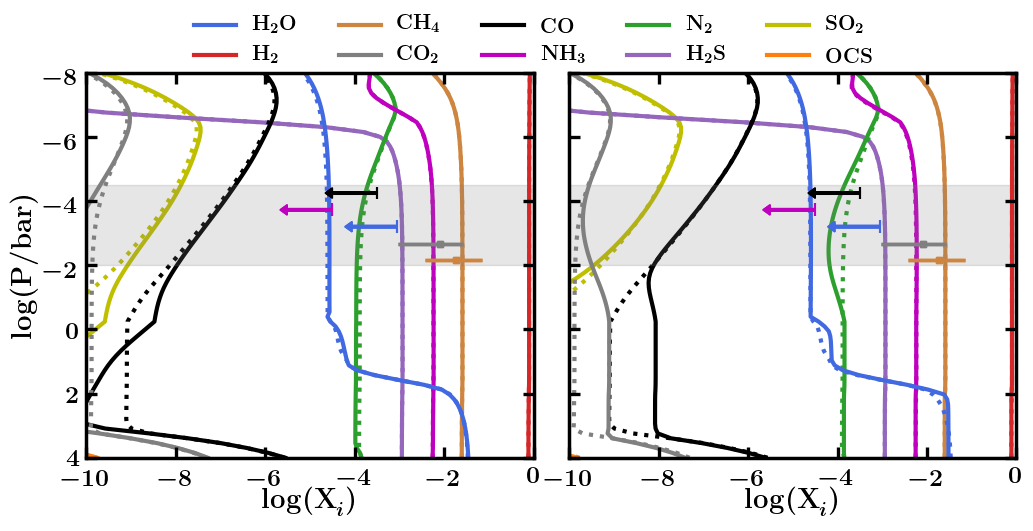}
    \caption{Effect of varying $K_\mathrm{zz}$ on the vertical mixing ratio profiles for several H-C-O-N-S molecular species for the C2 $P$-$T$ profile. Dotted lines show the profile from the left-hand side of Figure \ref{fig:Xprofiles}, for C2 with $30\%$ silicates, equivalent to $50\times$ solar elemental abundances with our canonical treatment of $K_\mathrm{zz}$, with a value of $10^6$~cm$^2$s$^{-1}$ in the deep atmosphere. Left: solid lines indicate the corresponding profiles with a lower $K_\mathrm{zz}=10^4$~cm$^2$s$^{-1}$ in the deep atmosphere. Right: solid lines indicate the corresponding profiles with a higher $K_\mathrm{zz}=10^8$~cm$^2$s$^{-1}$ in the deep atmosphere.}
    \label{fig:Xprofiles_Kzz}
\end{figure*}

Next we consider a range of $K_\mathrm{zz}$ values in the deep atmosphere, using the C2 $P$-$T$ profile. We vary $K_\mathrm{zz}$ at $P>0.5$~bar from $10^4$ to $10^8$~cm$^2$s$^{-1}$, with our canonical value at $10^6$~cm$^2$s$^{-1}$. The metallicity remains fixed at the canonical value of $50\times$ solar. As shown in Figure~\ref{fig:Xprofiles_Kzz}, both the higher and lower $K_\mathrm{zz}$ values negligibly affect the computed mixing ratios at observable pressures. Increasing $K_\mathrm{zz}$ shifts the quench point to higher (deeper) pressures, as shown by the mixing ratio profile for CO$_2$ in the right-hand panel of Figure~\ref{fig:Xprofiles_Kzz}.

\begin{figure*}
    \centering
    \includegraphics[width=0.9\textwidth]{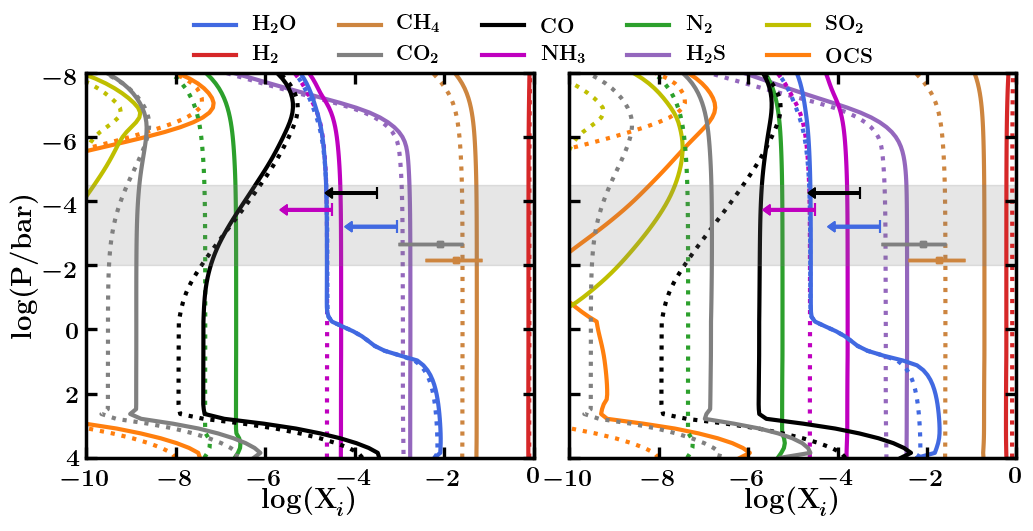}
    \caption{Effect of enhanced metallicity on the vertical mixing ratio profiles for several H-C-O-N-S molecular species with the C3 $P$-$T$ profile. Dotted lines show the profile from the right-hand side of Figure \ref{fig:Xprofiles}, for C3 with $30\%$ silicates, with N depletion due to the presence of magma. Left: solid lines indicate the corresponding profiles for 100$\times$solar metallicity, i.e. $2\times$ the respective elemental abundances given in Table~\ref{tab:extreme}. Right: solid lines indicate the corresponding profiles for 300$\times$solar metallicity, i.e. $6\times$ the respective elemental abundances in Table~\ref{tab:extreme}.}
    \label{fig:Xprofiles_metallicity_c3}
\end{figure*}

We now consider the hotter $P$-$T$ profile case with NH$_3$ depletion due to magma; this is the C3 profile with 30$\%$ silicates, as discussed above. The higher metallicities of 100$\times$ and 300$\times$ solar are implemented by proportionately enhancing the canonical elemental abundances for this case. These originally corresponded to 50$\times$ solar, hence we increase the relevant elemental abundances in Table \ref{tab:extreme} by factors of 2 and 6, respectively. The results are shown in Figure~\ref{fig:Xprofiles_metallicity_c3}. As for the C2 profile, the predicted CO$_2$ abundance remains significantly below the retrieved value in both cases, with the CO mixing ratio exceeding that of CO$_2$ throughout the atmosphere. The CH$_4$ abundance for $300\times$ solar metallicity is additionally too high compared to the retrieved abundance.

\begin{figure*}
    \centering
    \includegraphics[width=0.9\textwidth]{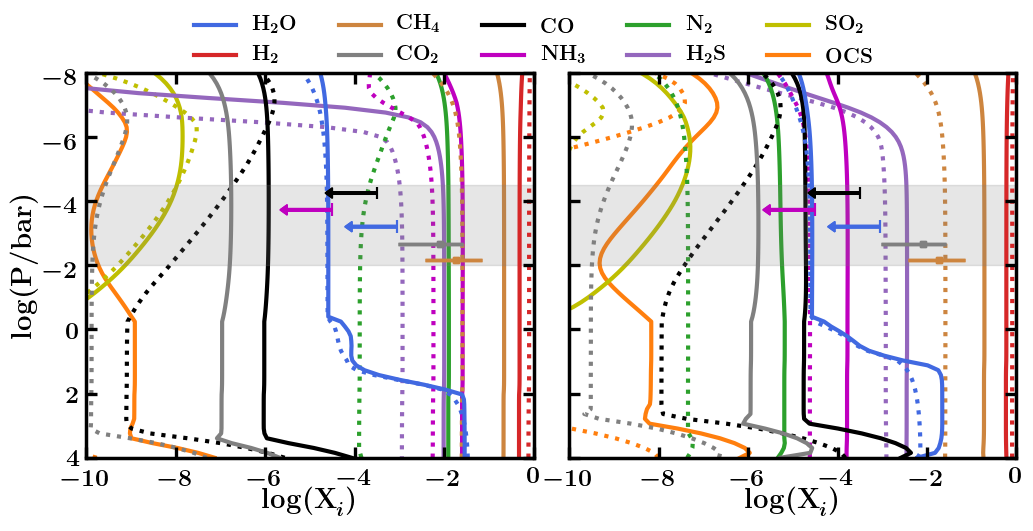}
    \caption{Effect of high metallicity and high $K_\mathrm{zz}$ on the vertical mixing ratio profiles for several H-C-O-N-S molecular species. Dotted lines show the canonical profiles from Figure \ref{fig:Xprofiles}. Left: solid lines indicate the profiles with the C2 $P$-$T$ profile with 300$\times$solar abundance and $K_\mathrm{zz}=10^8$ cm$^2$s$^{-1}$ in the deep atmosphere. Right: solid lines indicate the profiles with the C3 $P$-$T$ profile with 6$\times$ the elemental abundances from Table \ref{tab:extreme}, i.e. equivalent to 300$\times$solar abundance, and $K_\mathrm{zz}=10^8$ cm$^2$s$^{-1}$ in the deep atmosphere.}
    \label{fig:Xprofiles_highmet_highKzz}
\end{figure*}

As an end-member case, we consider each of the C2 and C3 profiles discussed above and adopt our extreme values of $300\times$solar metallicity and $K_\mathrm{zz}=10^8$ cm$^2$s$^{-1}$ in the deep atmosphere. The resulting vertical mixing ratio profiles are shown in Figure \ref{fig:Xprofiles_highmet_highKzz} along with the canonical cases. These end-member cases are similarly unable to match the retrieved CO$_2$ abundance constraints. A higher $K_\mathrm{zz}$ would further increase the abundances of both CO and CO$_2$, however CO remains more abundant than CO$_2$.

Thus far we have considered values of $25$~K and $50$~K for $T_\mathrm{int}$, corresponding the C2 and C3 profiles, respectively. Lastly, we explore the effect of increasing $T_\mathrm{int}$ to a higher value of $60$~K for completeness, as has been considered by other works for K2-18~b \citep[e.g.][]{Hu2021a}. We adopt the $P$-$T$ profile of \citet{Hu2021a} with $100\times$ solar metallicity, extrapolated to higher pressures ($1000$~bar) using an adiabat. We consider two cases: 1) $100\times$ solar metallicity with depletion (i.e. twice the C3 $30\%$ silicates abundances from Table~\ref{tab:extreme}) and our canonical $K_\mathrm{zz}$ treatment, and 2) a high $K_\mathrm{zz}=10^8$ cm$^2$s$^{-1}$ and a high metallicity of $300\times$ solar (i.e. $6\times$ the C3 $30\%$ silicates abundances). With the canonical $K_\mathrm{zz}$, we find that the computed CO abundance exceeds the retrieved upper limit, while the computed CO$_2$ abundance remains significantly lower than the retrieved abundance. The retrieved CH$_4$ abundance and NH$_3$ upper limits can be explained by this model. For the high $K_\mathrm{zz}$ and high metallicity case, the computed CO abundance similarly exceeds the retrieved abundance. In this case the retrieved CO$_2$ abundance can be explained by the model. However, the computed CH$_4$ abundance exceeds the retrieved value. Due to the higher temperatures in this $P$-$T$ profile, the H$_2$O abundance exceeds the retrieved value for both cases of metallicity and $K_\mathrm{zz}$ considered.

Overall, we have explored a wide parameter space for the atmospheric chemistry, considering a range of values for $K_\mathrm{zz}$, metallicity and $T_\mathrm{int}$. In this exploration, we do not find a case resulting in CO$_2$ $>$ CO that would satisfy the retrieved atmospheric abundance constraints for K2-18~b \citep{madhusudhan_carbon-bearing_2023}.

\begin{figure*}
    \centering
    \includegraphics[width=0.9\textwidth]{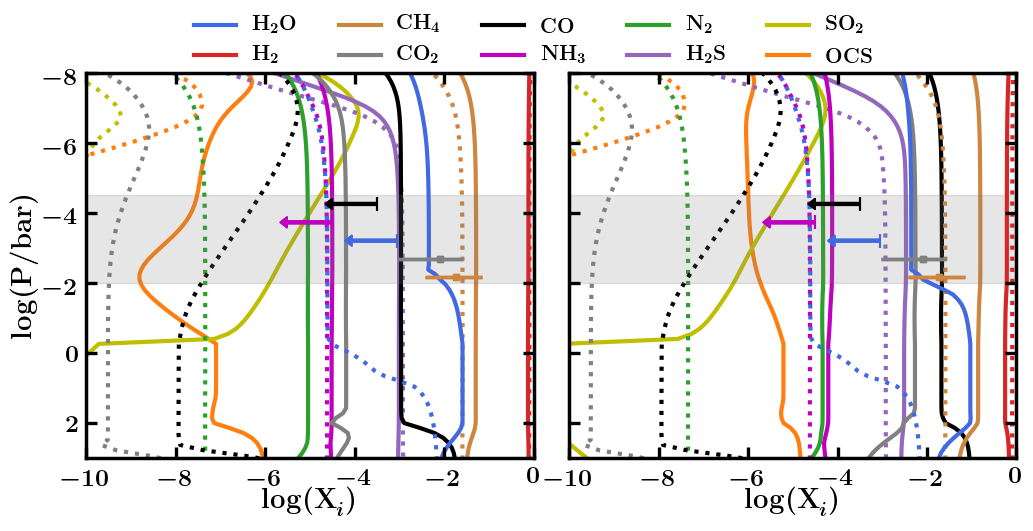}
    \caption{Effect of higher $T_\mathrm{int}$ on the vertical mixing ratio profiles for several H-C-O-N-S molecular species. We adopt the $P$-$T$ profile from \citet{Hu2021a} with $T_\mathrm{int}$ of $60$~K and $100\times$ solar metallicity, extrapolated to $1000$~bar. Dotted lines show the canonical profiles using the hotter C3 profile from Figure \ref{fig:Xprofiles}. Left: solid lines indicate profiles assuming $100\times$ solar metallicity including N depletion, adopting $K_\mathrm{zz}=10^6$ cm$^2$s$^{-1}$ in the deep atmosphere, for the \citet{Hu2021a} $P$-$T$ profile. Right: solid lines indicate profiles assuming $300\times$ solar metallicity including N depletion, adopting $K_\mathrm{zz}=10^8$ cm$^2$s$^{-1}$ in the deep atmosphere, for the \citet{Hu2021a} $P$-$T$ profile.}
    \label{fig:Xprofiles_Hu}
\end{figure*}

\end{document}